\documentclass[apj]{emulateapj}

\usepackage{graphicx}

\slugcomment{Accepted for publication in ApJ}

\shorttitle{A double plateau in the transient SN~2011A.}
\shortauthors{de Jaeger et al.}

\begin{document}

\title{SN~2011A: an low--luminosity interacting transient, with a double plateau and strong sodium absorption\altaffilmark{1}}

\altaffiltext{1}{This paper includes data obtained with the 6.5--m Magellan Telescopes and du Pont telescope; the Gemini--North Telescope, Mauna Kea, USA (Gemini Program GN--2010B--Q67, PI: Stritzinger); the PROMPT telescopes at Cerro Tololo Inter--American Observatory in Chile; with the Liverpool Telescope operated on the island of La Palma by Liverpool John Moores University in the Spanish Observatorio del Roque de los Muchachos of the Instituto de Astrofisica de Canarias with financial support from the UK Science and Technology Facilities Council; based on observations made with the Nordic Optical Telescope, operated by the Nordic Optical Telescope Scientific Association at the Observatorio del Roque de los Muchachos, La Palma, Spain, of the Instituto de Astrofisica de Canarias; the NTT from ESO Science Archive Facility under allocations 184.D--1151 and 184.D--1140 (PI: Benetti. S), at the Centro Astronómico Hispano Alemán (CAHA) at Calar Alto, operated jointly by the Max--Planck Institut f\"ur Astronomie and the Instituto de Astrofísica de Andalucía (CSIC), on observations collected at Asiago Observatory and the Southern Astrophysical Research (SOAR) telescope, which is a joint project of the Minist\'{e}rio da Ci\^{e}ncia, Tecnologia, e Inova\c{c}\~{a}o (MCTI) da Rep\'{u}blica Federativa do Brasil, the U.S. National Optical Astronomy Observatory (NOAO), the University of North Carolina at Chapel Hill (UNC), and Michigan State University (MSU).}

\author{T. \rm{de} Jaeger$^{2,3*}$, J. P. Anderson$^{4}$, G. Pignata$^{5,2}$, M. Hamuy$^{3,2}$, E. Kankare$^{6}$, M. D. Stritzinger$^{7}$, S. Benetti$^{8}$, F. Bufano$^{2,5}$, N. Elias-Rosa$^{8,9}$, G. Folatelli$^{10,11}$, F. F\"orster$^{2}$, S. Gonz\'alez-Gait\'an$^{2,3}$, C.P. Guti\'errez$^{2,3,4}$, C. Inserra$^{6}$, R. Kotak$^{6}$, P. Lira$^{3}$, N. Morrell$^{12}$, F. Taddia$^{13}$, L. Tomasella$^{8}$}

\affil{%
  (2) Millennium Institute of Astrophysics, Casilla 36--D, Santiago, Chile\\
  (3) Departamento de Astronom\'ia -- Universidad de Chile, Camino el Observatorio 1515, Santiago, Chile.\\
  (4) European Southern Observatory, Alonso de Cordova 3107, Vitacura, Santiago, Chile\\
  (5) Departamento de Ciencias F\'isicas -- Universidad Andres Bello, Avda. Rep\'ublica 252, Santiago, Chile.\\
  (6) Astrophysics Research Center, School of Mathematics and Physics, Queens University Belfast, Belfast, BT7 1NN, UK.\\
  (7) Department of Physics and Astronomy, Aarhus University, Ny Munkegade 120, DK--8000 Aarhus C, Denmark.\\
  (8) INAF--Osservatorio Astronomico di Padova, vicolo dell'Osservatorio 5, I--35122, Padova, Italia.\\
  (9) Institut de Ci\`encies de l'Espai (IEEC-CSIC), Facultat de Ci\`encies, Campus UAB, 08193 Bellaterra, Spain.\\
  (10) Instituto de Astrofísica de La Plata, CONICET, Paseo del Bosque S/N, B1900FWA, La Plata, Argentina.\\
  (11) Kavli Institute for the Physics and Mathematics of the Universe (WPI), Todai Institutes for Advanced Study, the University of Tokyo, Kashiwa, Japan 277--8583 (Kavli IPMU, WPI).\\
  (12) Las Campanas Observatory, Carnegie Observatories, Casilla 601, La Serena, Chile.\\
  (13) The Oskar Klein Centre, Department of Astronomy, Stockholm University, AlbaNova, 10691 Stockholm, Sweden.\\
}

\email{*dthomas@das.uchile.cl}

\begin{abstract}

We present optical photometry and spectroscopy of the optical transient SN~2011A. Our data spans 140 days after discovery including $BVRIu'g'r'i'z'$  photometry and 11 epochs of optical spectroscopy. Originally classified as a type IIn supernova (SN~IIn) due to the presence of narrow H$\alpha$ emission, this object shows exceptional characteristics. Firstly, the light curve shows a double plateau; a property only observed before in the impostor SN~1997bs. Secondly, SN~2011A has a very low luminosity ($M_{V}=-15.72$), placing it between normal luminous SNe~IIn and SN~impostors. Thirdly, SN~2011A shows low velocity and high equivalent width absorption close to the sodium doublet, which increases with time and is most likely of circumstellar origin. This evolution is also accompanied by a change of line profile; when the absorption becomes stronger, a P--Cygni profile appears. We discuss SN~2011A in the context of interacting SNe~IIn and SN~impostors, which appears to confirm the uniqueness of this transient. While we favour an impostor origin for SN~2011A, we highlight the difficulty in differentiating between terminal and non--terminal interacting transients.
\end{abstract}


\section{Introduction}

It is generally accepted that the majority of massive stars with zero--age main--sequence mass
$\geq$ 8 ${\rm M}_{\odot}$ \citep{smart09b} end their lives as core--collapse supernovae (CC SNe), with the possibility that some directly form black holes with no visible supernovae \citep{smartt15}. CC SNe are classified in two groups according to the absence (SNe~Ib/c) or presence (SNe~II) of \ion{H}{1} lines \citep{filippenko97}.
The ejected material from a SN explosion is sometimes observed to interact with surrounding circumstellar material (CSM), related to progenitor mass--loss episodes prior to explosion \citep{che81,fra82}.\\
\indent
When the CSM is sufficiently dense, strong CSM--ejecta interaction can begin shortly after explosion; 
this is often observed as a type IIn supernova (SN~IIn henceforth) \citep{sch90,chu94}. 
These objects are characterized by prevalent blue continua and narrow emission
lines in their spectra superimposed on broader emission profiles. The H$\alpha$ emission typically dominates the
spectrum and often has multiple components consisting of (e.g., \citealt{sal98}): 

\begin{enumerate}
\item {narrow emission {$(v_{FWHM}$ ${\sim}$  few hundred km s$^{-1}$)} formed by the photoionization 
of the high density pre--existing CSM from the prompt emission of the SN. 
The high density has furthermore been attributed to a high mass--loss rate \citep{chu04}.} 
\item {intermediate emission $(v_{FWHM}$ ${\sim}$ few thousand km s$^{-1}$) possibly produced 
from the shock interaction of the SN blast within a dense shell of clumpy CSM or with a dense equatorial wind \citep{chu94}.}
\item {broad emission ($v_{FWHM}$ $\geq 10{^4}$ km s$^{-1}$) most likely due to SN ejecta 
as a result of multiple scattering by thermal electrons in the opaque CS gas \citep{chu01}.
The presence of this broad H$\alpha$ emission without broad P--Cygni absorption is usually 
considered to be caused by a dense wind \citep{chu90}.}
\end{enumerate}

Measured parameters of these components can provide valuable information on progenitor 
properties and its late time evolution (such as the mass--loss rate or wind velocity), 
through the observed properties of their CSM.\\ 

Observations of SNe IIn published to date (see e.g., \citealt{kie12,tad13,zhang12}) 
have allowed researchers to probe the immediate environment of these explosions and
point towards significant diversity among various SNe~IIn. The high mass--loss rates calculated from the observations of various SNe~IIn (e.g., \citealt{kie12,tad13,moriya14}) have prompted several authors to assign Luminous Blue Variable stars (LBVs) in several cases as immediate progenitors of SNe~IIn. LBVs, as defined by \citet{hum94}, are some of the most
luminous stars, with episodic and violent mass--loss events with mass--loss rates $\geq {10^{-3}M_{\sun}}$ yr$^{-1}$.
Classic examples of LBVs are Eta Carinae and P--Cygni. The direct detection of the extremely luminous progenitor of SN~2005gl
\citep{gal09} also supports this possibility. However, in stellar evolutionary models, the LBV phase follows a blue giant star state \citep{crowther2007,gra08}, most likely with mass loss driven by the 
bi--stability jump \citep{vin11}, and it is then followed by a H--poor WN star 
(nitrogen rich Wolf--Rayet). In this scenario the LBV star loses its H envelope, 
becomes a Wolf--Rayet star and then explodes as a SN. 
Thus, LBVs have been placed as a post--main sequence, 
but not a final pre--SN phase \citep{sch92,lan93,sto96,mae05,mae08,dwa10}, 
although see \citet{groh2013} for recent modeling suggesting that some stars 
could indeed end their lives during this phase.\\

There have been a number of cases where transients have originally
been defined as SNe~IIn, but a link to a definitive SN event (i.e. the terminal 
stage of a star's life) is questionable. Such events have been defined as SN
``impostors'' \citep{van00,maund06} which are believed to be luminous non--terminal 
eruptions of massive stars instead of SNe. There are multiple cases of such events e.g., 
the on--going transient SN~2009ip (\citealt{pas12,pri12,mau12,fraser13,margutti13,smith14b}).
SN~impostors are generally claimed to be LBV transients with peak luminosities lower than SNe ($M\geq -14$), such as SN~1997bs ($M_{V}\sim -13.8$; \citealt{van00}), SN~2000ch ($M_{R}\sim -12.8$; \citealt{wagner2004,pastorello2010}), SN~2002kg ($M_{V}\sim -9.6$; \citealt{van06,maund06,weis05}), SN~2007sv ($M_{R}\sim -14.2$; \citealt{tartaglia14}), SN 2008S ($M_{R}\sim -13.9$; \citealt{kochanek11,smi09}, and 2008 NGC 300-OT ($M_{V}\sim -12~to~-13$; a\citealt{bond09} (see also the compilation paper by \citealt{smith11}). However, unambiguous evidence that all transients classified as ``impostors'' are actually non--terminal events is lacking. Indeed for the first transient identified as a SN~impostor, SN~1997bs, it has been claimed that no surviving star is present \citep{li2002,adams15}. \citet{adams15} have argued that the scenario where the surviving star is obscured by dust created in the eruption is excluded.\\

Given that SN~impostors and SNe~IIn show similar spectroscopic features, such as the presence of 
narrow emission lines (in particular \ion{H}{1} lines), the differentiation based solely on spectral analysis is rendered difficult, and constraining the nature of their progenitors (e.g., ZAMS mass) has been problematic.
Indeed, the detection of narrow emission lines is merely evidence for high density material close to the SN
and it appears that this high density CSM can be produced by a number of different progenitor scenarios. Indeed, it has even been observed in SNe~Ia (e.g., SN~2002ic \citealt{hamuy03}, SN~2011kx \citealt{dil12}).\\

While observations of individual SNe IIn have been used to argue for LBV
progenitors or LBV transients, statistical investigations of their environments within host
galaxies have pointed to lower mass progenitors. \citet{and12} and \citet{habergham14} found
a low correlation between SNe~IIn and host--galaxy \ion{H}{2} regions. It was
found that SNe~IIn show a similar degree of association with star formation (SF) 
as SNe~IIP, which are thought to arise from stars at the lower end of the CC SNe scale; 
with masses between 8--16 ${\rm M}_{\odot}$ \citep{smart09b}. Therefore, these results 
suggest that a significant fraction of SNe~IIn arise from relatively low mass progenitors. 
This is somewhat surprising given that the LBV phase is thought to be associated only 
with very massive stars ($\sim$ 30--80 ${\rm M}_{\odot}$). Contrary to the above, we note that recent results suggest that a significant fraction of SNe~IIn could arise from electron capture SNe \citep{smi13} with AGB progenitor stars (\citealt{prieto09,bot2009}). Also, \citet{smith15} argue that LBV stars are the product of binary kicks, due to their isolated environments.\\

In this paper we present optical photometry and spectroscopy of the
transient SN~2011A, originally classified as a SN~IIn \citep{str11}. 
Given the properties of the transient we discuss its characteristics 
in comparison to other SN~IIn and SN~impostors with the aim of defining 
the nature of the transient and constraining its properties. 
Due to the fact that it is unclear whether SN~2011A was a true SN, 
we refer to it as ``transient'' throughout the paper.\\

The paper is organised as follows. Section 2 contains a description of the
observations and in Section 3 we discuss our results, first analysing 
light-- and color--curves and then using spectroscopy to derive physical parameters for the transient. 
In Section 4 we discuss our results and we conclude with a summary in Section 5.
 
\section{Observations} 

\subsection{Discovery}

SN~2011A  was discovered on an unfiltered image (apparent magnitude $\sim$ 16.9) on 
2011 January 2.30 UT with the 0.41--m `PROMPT 4' telescope located at Cerro Tololo
\citep{pig11} on behalf of the CHASE project \citep{pig09}. The transient
was located at $\alpha = 13\fh 01\fm 01\fs19 \pm 0\fs2 $, $\delta
= -14^{\circ}31{\farcm}34{\farcs}8 \pm  0{\farcs}2$  (J2000.0), $21{\farcs}3$
east and $46{\farcs}5$  south of center of the galaxy NGC 4902 at z=0.008916 (\citealt{theureau07}, see NED\footnote{http://ned.ipac.caltech.edu/}). NGC 4902 is a luminous Sb galaxy ($M_{B} = -21.40$). \citet{str11} classified the object to be a SNe~IIn due
to the presence of prevalent narrow Balmer lines in the spectrum. 
The most recent pre--explosion non detection on archival images was dated 2010 July 13.05 (limiting magnitude
18.5) providing extremely weak constraints on any explosion time or previous
variability. All observations are presented with respect to the discovery date: 
2011 January 2.30 UT. In Fig.~1 the field of the transient is displayed. 
\begin{figure}
\epsscale{1.18}
\plotone{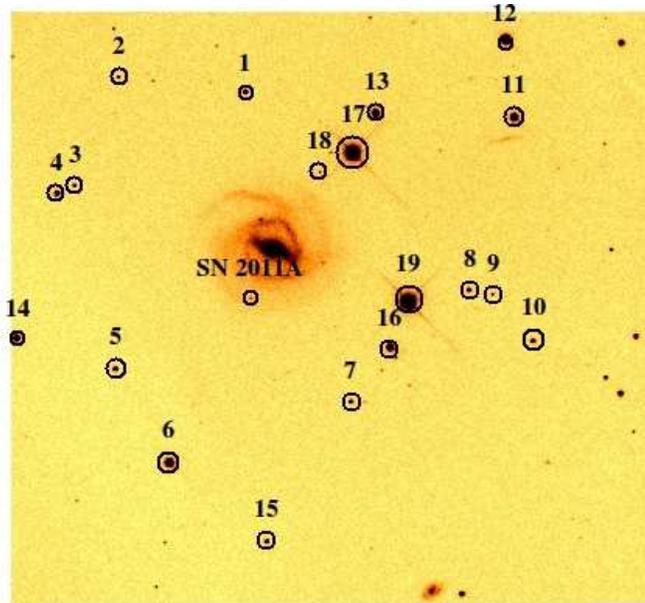}
\caption {The field of SN~2011A showing the location of sequence stars (designated numerically) listed in Table 1. The location of SN~2011A is marked by a grey circle. North is up and East is left. The image is $10\arcmin \times 10\arcmin$  and was taken with PROMPT--5 on 2011 January 23 in $r'$ band.}
\label{Figure.1}
\end{figure}

\subsection{Photometry}

Optical photometry was obtained with the Panchromatic Robotic Optical
Monitoring  and Polarimetry Telescopes (PROMPT; \citealt{reichart05}) at the Cerro Tololo Inter--American Observatory, RATCam 
mounted on the 2.0m Liverpool Telescope \citep{steele2004} (LT) at the Observatorio del Roque de  los Muchachos on the Canary Island of La Palma, with EFOSC2 \citep{buzzoni1984} mounted on the 3.6m New Technology Telescope (NTT) at the La Silla Observatory, the 2.2m telescope at the Calar Alto (CA) Observatory with the optical
imager/spectrograph CAFOS, and from the imager AFOSC mounted at the 1.82m
reflector  telescope at the Asiago Observatory.  
All images were analysed after automated reductions
(bias--, flatfield--correction and astrometric solution) had been applied and all the science images were host--galaxy subtracted.\\

Photometry of the transient was computed relative to a sequence of 19 
stars in the field of NGC 4902, which we calibrate using 
catalogs from \citet{lan92,lan07} for $B, V, R , I$ bands and \citet{smithja2002} 
for $u', g', r', i', z'$ bands. Instrument--specific colours terms were 
derived using several epochs of standard field observations. 
The resulting $B, V, R, I, u', g', r', i', z'$ magnitudes
of the local sequence stars are reported in Table 1, which 
correspond to the averages from n photometric nights (n=2 for the $z'$ filter
and n$\ge$3 for the remaining filters), and the uncertainties are the standard
deviations from the mean.\\
 
Differential photometry of the transient was done relative to the local sequence.
To obtain instrumental magnitudes, we used the IRAF\footnote{IRAF (Image
Reduction  and Analysis Facility) is distributed by the National Optical
Astronomy Observatories, which are operated by AURA, Inc., under cooperative
agreement with the National Science Foundation.} package \textit{SNOOPY}\footnote{SNOOPY, originally presented in \citet{patat1996}, has been implemented in IRAF by E. Cappellaro.}. 
Photometry for SN~2011A is presented in Table 2 and the resulting light curve is displayed in Fig.~2.
Errors come from the point spread function (PSF) fit and the local sequence magnitudes.\\

\begin{table*} 
\tiny
\begin{center}
\caption{$B$, $V$, $R$, $I$, {\rm $u'$}, {\rm $g'$}, {\rm $r'$}, {\rm $i'$}, {\rm $z'$} local sequence star magnitudes.\label{Table 1.}}
\begin{tabular}{lccccccccccc}
\tableline\tableline
STAR  &R.A &Decl &$B$ &$V$ &$R$ &$I$ & $u'$ &$g'$ &$r'$ & $i'$ & $z'$\\
\tableline
1 &13:01:01.523  &$-$14:28:21.35 &15.78$\pm$0.02   &15.20$\pm$0.04  &14.83$\pm$0.03 &14.46$\pm$0.05 &16.65$\pm$0.02  &15.45$\pm$0.02 &15.05$\pm$0.02 &14.90$\pm$0.04 &14.91$\pm$0.01 \\
2 &13:01:09.706 &$-$14:28:07.32 &17.69$\pm$0.05   &17.06$\pm$0.04  &16.69$\pm$0.03 &16.32$\pm$0.05 &18.57$\pm$0.01 &17.31$\pm$0.03  &16.91$\pm$0.01 &16.74$\pm$0.03 &16.76$\pm$0.02  \\
3 &13:01:12.549 &$-$14:29:48.59 &17.20$\pm$0.02   &16.41$\pm$0.02  &15.95$\pm$0.03 &15.51$\pm$0.04 &18.35$\pm$0.09 &16.75$\pm$0.02  &16.19$\pm$0.02 &15.96$\pm$0.03 &15.94$\pm$0.01 \\
4 &13:01:13.685  &$-$14:29:56.09 &15.68$\pm$0.02   &14.70$\pm$0.03  &14.12$\pm$0.03 &13.65$\pm$0.04 &17.66$\pm$0.05 &15.16$\pm$0.04  &14.38$\pm$0.01 &14.11$\pm$0.03 &14.06$\pm$0.01  \\
5  &13:01:09.906 &$-$14:32:40.66 &15.96$\pm$0.01  &15.10$\pm$0.03  &14.61$\pm$0.02 &14.21$\pm$0.41 &17.60$\pm$0.03 &15.50$\pm$0.04  &14.86$\pm$0.01 &14.65$\pm$0.03 &14.65$\pm$0.01 \\
6  &13:01:06.445 &$-$14:34:08.13 &13.42$\pm$0.01   &12.43$\pm$0.03  &11.87$\pm$0.03 &11.36$\pm$0.03 &15.19$\pm$0.02 &12.88$\pm$0.04  &12.14$\pm$0.01 &11.83$\pm$0.03 &11.74$\pm$0.01  \\
7  &13:00:54.730 &$-$14:33:11.10 &17.33$\pm$0.02  &16.17$\pm$0.02  &15.44$\pm$0.02 &14.84$\pm$0.04 &19.52$\pm$0.10 &16.74$\pm$0.04  &15.72$\pm$0.01 &15.32$\pm$0.03 &15.19$\pm$0.01  \\
8 &13:00:47.126 &$-$14:31:26.73 &16.46$\pm$0.01  &15.55$\pm$0.02  &14.98$\pm$0.02 &14.48$\pm$0.04 &18.27$\pm$0.04  &15.99$\pm$0.03 &15.23$\pm$0.02 &14.94$\pm$0.03 &14.87$\pm$0.02 \\
9  &13:00:45.583 &$-$14:31:31.08 &17.99$\pm$0.03  &17.36$\pm$0.01  &17.00$\pm$0.03 &16.66$\pm$0.04 &18.72$\pm$0.02 &17.62$\pm$0.02  &17.24$\pm$0.02 &17.08$\pm$0.05 &17.08$\pm$0.01  \\
10  &13:00:42.987  &$-$14:32:14.39 &16.06$\pm$0.01   &15.42$\pm$0.03  &15.03$\pm$0.02 &14.64$\pm$0.04 &16.95$\pm$0.02 &15.69$\pm$0.02  &15.26$\pm$0.02 &15.08$\pm$0.03 &15.07$\pm$0.01 \\
11  &13:00:44.202 &$-$14:28:44.77 &13.21$\pm$0.03   &12.73$\pm$0.03  &12.44$\pm$0.03 &12.17$\pm$0.04 &14.21$\pm$0.02 &12.92$\pm$0.03  &12.65$\pm$0.02 &12.58$\pm$0.03 &12.64$\pm$0.01  \\
12  &13:00:44.752  &$-$14:27:31.79 &12.69$\pm$0.04   &12.09$\pm$0.03  &11.72$\pm$0.02 &11.38$\pm$0.03 &13.69$\pm$0.03 &12.36$\pm$0.01  &11.94$\pm$0.02 &11.81$\pm$0.04 &11.83$\pm$0.01 \\
13  &13:00:53.129 &$-$14:28:40.67 &13.90$\pm$0.02   &13.11$\pm$0.03  &12.65$\pm$0.03 &12.20$\pm$0.04 &15.15$\pm$0.02 &13.46$\pm$0.02  &12.90$\pm$0.02 &12.66$\pm$0.03 &12.61$\pm$0.01  \\
14  &13:01:16.297  &$-$14:32:11.47 &14.07$\pm$0.01   &13.48$\pm$0.03  &13.15$\pm$0.03 &12.83$\pm$0.04 &14.99$\pm$0.02 &13.72$\pm$0.02  &13.37$\pm$0.01 &13.25$\pm$0.03 &13.30$\pm$0.01  \\
15 &13:01:00.164  &$-$14:35:22.03   &15.98$\pm$0.01   &15.33$\pm$0.02  &14.93$\pm$0.02 &12.83$\pm$0.04 &16.74$\pm$0.01  &15.60$\pm$0.02 &15.16$\pm$0.02 &14.97$\pm$0.03 &14.96$\pm$0.01 \\
16  &13:00:52.187  &$-$14:32:20.12 &13.95$\pm$0.01   &13.16$\pm$0.02  &12.70$\pm$0.03 &12.31$\pm$0.04 &15.35$\pm$0.02 &13.52$\pm$0.02  &12.94$\pm$0.02 &12.75$\pm$0.03 &12.72$\pm$0.01  \\
17 &13:00:54.651  &$-$14:29:18.73  &11.16$\pm$0.03  &10.38$\pm$0.02 &$\cdots$  &$\cdots$ &$\cdots$ &10.65$\pm$0.03 &10.57$\pm$0.04  &10.39$\pm$0.03 &9.74$\pm$0.06  \\
18  &13:00:56.730  &$-$14:29:37.00 &19.93$\pm$0.05   &18.61$\pm$0.03  &17.74$\pm$0.01 &17.10$\pm$0.06 &20.60$\pm$1.56 &19.19$\pm$0.10  &18.09$\pm$0.01 &17.57$\pm$0.03 &17.35$\pm$0.02 \\
19 &13:00:51.075  &$-$14:31:37.80  &10.84$\pm$0.02  &10.33$\pm$0.03 &$\cdots$  &$\cdots$  &$\cdots$ &10.48$\pm$0.06 &10.58$\pm$0.05   &10.41$\pm$0.01 &9.66$\pm$0.05\\  
\tableline
\end{tabular}
\\
\end{center}
\begin{footnotesize}
 \hspace{2cm}
Note. -- Errors correspond to the standard deviation from the mean.{\let\thefootnote\relax}
\end{footnotesize}
\end{table*}

\begin{table*} 
\tiny
\begin{center}
\caption{Optical photometry of SN~2011A.
\label{Table 2.}}
\tabcolsep2pt
\begin{tabular}{l c c c c c c c c c c c}
\tableline\tableline 
JD$-2450000+$ &Epoch\tablenotemark{a} &$B$ &$V$ &$R$ &$I$ &$u'$ &$g'$ &$r'$ &$i'$ &$z'$ &Telescope/Instrument\\
\tableline
5566.3  &2.5  &18.11$\pm$0.02 &17.86$\pm$0.02 &17.51$\pm$0.02 &17.38$\pm$0.02  &$\cdots$ &$\cdots$ &$\cdots$ &$\cdots$ &$\cdots$ &PROMPT/Apogée   \\
5567.3  &3.5  &18.19$\pm$0.04 &17.86$\pm$0.02 &17.53$\pm$0.01 &17.26$\pm$0.05  &$\cdots$ &$\cdots$ &$\cdots$ &$\cdots$ &$\cdots$ &PROMPT/Apogée\\
5569.3  &5.5  &18.18$\pm$0.03 &17.85$\pm$0.02 &17.49$\pm$0.04 &17.40$\pm$0.02  &$\cdots$ &$\cdots$ &$\cdots$ &$\cdots$ &$\cdots$ &PROMPT/Apogée   \\
5570.3  &6.5  &$\cdots$ &17.87$\pm$0.02 &17.50$\pm$0.04 &17.45$\pm$0.02 &$\cdots$ &$\cdots$ &$\cdots$ &$\cdots$ &$\cdots$ &PROMPT/Apogée   \\
5573.3  &9.5  &18.23$\pm$0.03 &17.87$\pm$0.02 &17.54$\pm$0.04 &17.41$\pm$0.02 &$\cdots$ &18.05$\pm$0.02 &17.82$\pm$0.02 &17.89$\pm$0.03 &17.78$\pm$0.09 &PROMPT/Apogée   \\
5573.6  &9.8  &18.23$\pm$0.04 &17.90$\pm$0.03 &$\cdots$ &$\cdots$ &$\cdots$ &$\cdots$ &$\cdots$ &$\cdots$ &$\cdots$ &LT/RATCam   \\
5575.3  &11.5  &$\cdots$ &17.89$\pm$0.02 &17.56$\pm$0.06 &17.33$\pm$0.04 &$\cdots$ &$\cdots$ &$\cdots$ &$\cdots$ &$\cdots$ &PROMPT/Apogée   \\
5575.7  &11.9  &18.22$\pm$0.06 &$\cdots$ &$\cdots$ &$\cdots$ &$\cdots$ &19.04$\pm$0.08 &$\cdots$ &$\cdots$ &$\cdots$ &LT/RATCam   \\
5576.2  &12.4  &18.26$\pm$0.02 &17.98$\pm$0.03 &17.65$\pm$0.04 &17.41$\pm$0.05 &$\cdots$ &$\cdots$ &$\cdots$ &$\cdots$ &$\cdots$ &CA--2.2/CAFOS   \\
5577.3  &13.5  &$\cdots$ &17.89$\pm$0.02 &17.56$\pm$0.06 &17.30$\pm$0.04 &$\cdots$ &$\cdots$ &$\cdots$ &$\cdots$ &$\cdots$ &PROMPT/Apogée   \\
5579.6  &15.8  &18.22$\pm$0.03 &17.96$\pm$0.02 &$\cdots$ &$\cdots$ &18.77$\pm$0.08 &18.13$\pm$0.03 &17.92$\pm$0.02 &17.91$\pm$0.03 &18.00$\pm$0.04 &LT/RATCam   \\
5581.3  &17.5  &18.35$\pm$0.03 &17.95$\pm$0.02 &17.57$\pm$0.04 &17.55$\pm$0.03 &$\cdots$ &$\cdots$ &$\cdots$ &$\cdots$ &$\cdots$ &PROMPT/Apogée   \\
5583.3  &19.5  &18.50$\pm$0.04 &18.02$\pm$0.03 &17.62$\pm$0.06 &17.45$\pm$0.05 &$\cdots$ &$\cdots$ &17.97$\pm$0.02 &17.68$\pm$0.06 &$\cdots$ &PROMPT/Apogée  \\
5584.3  &20.5  &$\cdots$ &18.10$\pm$0.03 &17.70$\pm$0.04 &17.51$\pm$0.03 &$\cdots$ &$\cdots$ &18.02$\pm$0.02 &18.05$\pm$0.03 &18.02$\pm$0.05 &PROMPT/Apogée  \\
5585.7  &21.9  &18.41$\pm$0.03 &18.20$\pm$0.12 &17.91$\pm$0.12 &$\cdots$ &$\cdots$ &$\cdots$ &$\cdots$ &$\cdots$ &$\cdots$ &NTT/EFOSC2   \\
5588.3  &24.5  &18.70$\pm$0.04 &18.24$\pm$0.03 &17.90$\pm$0.03 &17.67$\pm$0.13 &$\cdots$ &18.43$\pm$0.04 &18.04$\pm$0.04 &18.1$\pm$0.080 &18.42$\pm$0.10 &PROMPT/Apogée   \\
5591.3  &27.5  &$\cdots$ &$\cdots$ &17.97$\pm$0.05 &17.65$\pm$0.04 &$\cdots$ &$\cdots$ &$\cdots$ &$\cdots$ &$\cdots$ &PROMPT/Apogée \\
5600.3  &36.5  &$\cdots$ &18.27$\pm$0.02 &17.98$\pm$0.02 &17.62$\pm$0.03 &$\cdots$ &$\cdots$ &$\cdots$ &$\cdots$ &$\cdots$ &PROMPT/Apogée   \\
5600.7  &36.9  &18.81$\pm$0.02 &18.29$\pm$0.02 &$\cdots$ &$\cdots$ &19.65$\pm$0.06 &18.51$\pm$0.02 &18.20$\pm$0.02 &18.20$\pm$0.01 &18.14$\pm$0.04 &LT/RATCam \\
5602.3  &38.5  &$\cdots$ &$\cdots$ &$\cdots$ &$\cdots$ &$\cdots$ &$\cdots$ &18.20$\pm$0.02 &18.17$\pm$0.02 &18.15$\pm$0.03 &PROMPT/Apogée  \\
5606.3  &42.5  &18.99$\pm$0.04. &18.29$\pm$0.02 & &$\cdots$ &$\cdots$ &$\cdots$ &18.24$\pm$0.02 &18.11$\pm$0.02 &18.05$\pm$0.04 &PROMPT/Apogée  \\
5610.2  &46.4  &$\cdots$ &18.40$\pm$0.04 &18.06$\pm$0.03 &17.58$\pm$0.05 &$\cdots$ &$\cdots$ &17.98$\pm$0.07 &18.10$\pm$0.03 &18.11$\pm$0.01 &PROMPT/Apogée \\
5610.6  &46.8  &$\cdots$ &$\cdots$ &$\cdots$ &$\cdots$ &20.45$\pm$0.20 &18.80$\pm$0.06 &18.30$\pm$0.04 &18.22$\pm$0.03 &18.22$\pm$0.05 &LT/RATCam   \\
5613.2  &49.4  &$\cdots$ &$\cdots$ &$\cdots$ &$\cdots$ &$\cdots$ &$\cdots$ &18.24$\pm$0.06 &$\cdots$ &$\cdots$ &PROMPT/Apogée  \\
5615.3  &51.5  &19.39$\pm$0.05 &18.55$\pm$0.03 &18.02$\pm$0.06 &17.63$\pm$0.03 &$\cdots$ &$\cdots$ &$\cdots$ &$\cdots$ &$\cdots$ &PROMPT/Apogée   \\
5615.6  &51.8  &$\cdots$ &$\cdots$ &$\cdots$ &$\cdots$ &$\cdots$ &18.96$\pm$0.03 &18.37$\pm$0.02 &18.29$\pm$0.03 &18.14$\pm$0.03 &LT/RATCam   \\
5617.3  &53.5  &19.36$\pm$0.04 &18.55$\pm$0.01 &18.06$\pm$0.02 &17.80$\pm$0.03 &$\cdots$ &$\cdots$ &$\cdots$ &$\cdots$ &$\cdots$ &PROMPT/Apogée   \\
5621.6  &57.8  &$\cdots$ &$\cdots$ &$\cdots$ &$\cdots$ &$\cdots$ &19.16$\pm$0.02 &18.50$\pm$0.01 &18.28$\pm$0.02 &18.26$\pm$0.02 &LT/RATCam   \\
5623.2  &59.4  &19.53$\pm$0.01 &18.71$\pm$0.05 &18.08$\pm$0.04 &17.61$\pm$0.05 &$\cdots$ &$\cdots$ &18.25$\pm$0.05 &$\cdots$ &18.22$\pm$0.05 &PROMPT/Apogée  \\
5633.2  &69.4  &$\cdots$ &18.94$\pm$0.03 &18.07$\pm$0.06 &17.83$\pm$0.02 &$\cdots$ &$\cdots$ &$\cdots$ &$\cdots$ &$\cdots$ &PROMPT/Apogée  \\
5646.2  &82.4  &$\cdots$ &$\cdots$ &$\cdots$ &$\cdots$ &$\cdots$ &$\cdots$ &19.15$\pm$0.02 &18.69$\pm$0.02 &18.63$\pm$0.04 &PROMPT/Apogée  \\
5647.1  &83.3  &$\cdots$ &$\cdots$ &$\cdots$ &$\cdots$ &$\cdots$ &20.40$\pm$0.05 &18.91$\pm$0.04 &18.80$\pm$0.03 &$\cdots$ &PROMPT/Apogée  \\
5651.2  &87.4  &$\cdots$ &19.54$\pm$0.06 &$\cdots$ &18.44$\pm$0.05 &$\cdots$ &$\cdots$ &$\cdots$ &$\cdots$ &$\cdots$ &PROMPT/Apogée  \\
5651.5  &87.7  &$\cdots$ &$\cdots$ &$\cdots$ &$\cdots$ &$\cdots$ &20.43$\pm$0.09 &19.31$\pm$0.14 &18.87$\pm$0.33 &18.85$\pm$0.88 &LT/RATCam   \\
5653.1  &89.3  &$\cdots$ &19.59$\pm$0.06 &19.07$\pm$0.02 &$\cdots$ &$\cdots$ &$\cdots$ &$\cdots$ &$\cdots$ &$\cdots$ &PROMPT/Apogée \\
5661.4  &97.6  &$\cdots$ &20.20$\pm$0.22 &19.30$\pm$0.11 &18.79$\pm$0.20 &$\cdots$ &$\cdots$ &$\cdots$ &$\cdots$ &$\cdots$ &1.82m Reflector/AFOSC   \\
5666.2  &102.4  &$\cdots$ &$\cdots$ &$\cdots$ &$\cdots$ &$\cdots$ &$\cdots$ &19.51$\pm$0.12 &19.34$\pm$0.10 &$\cdots$ &PROMPT/Apogée \\
5677.1  &113.3  &$\cdots$ &$\cdots$ &19.54$\pm$0.06 &19.14$\pm$0.05 &$\cdots$ &$\cdots$ &$\cdots$ &$\cdots$ &$\cdots$ &PROMPT/Apogée \\
\tableline
\end{tabular}
\tablenotetext{1}{Days after discovery 2011 January 2.30 UT.}
\end{center}
\begin{footnotesize}
\let\thetablenote\relax
{Note. -- Errors correspond to one sigma uncertainties and come from the PSF fit and the local sequence magnitudes.}\\
\end{footnotesize}
\end{table*}
\begin{figure*}
\epsscale{1.40}
\plottwo{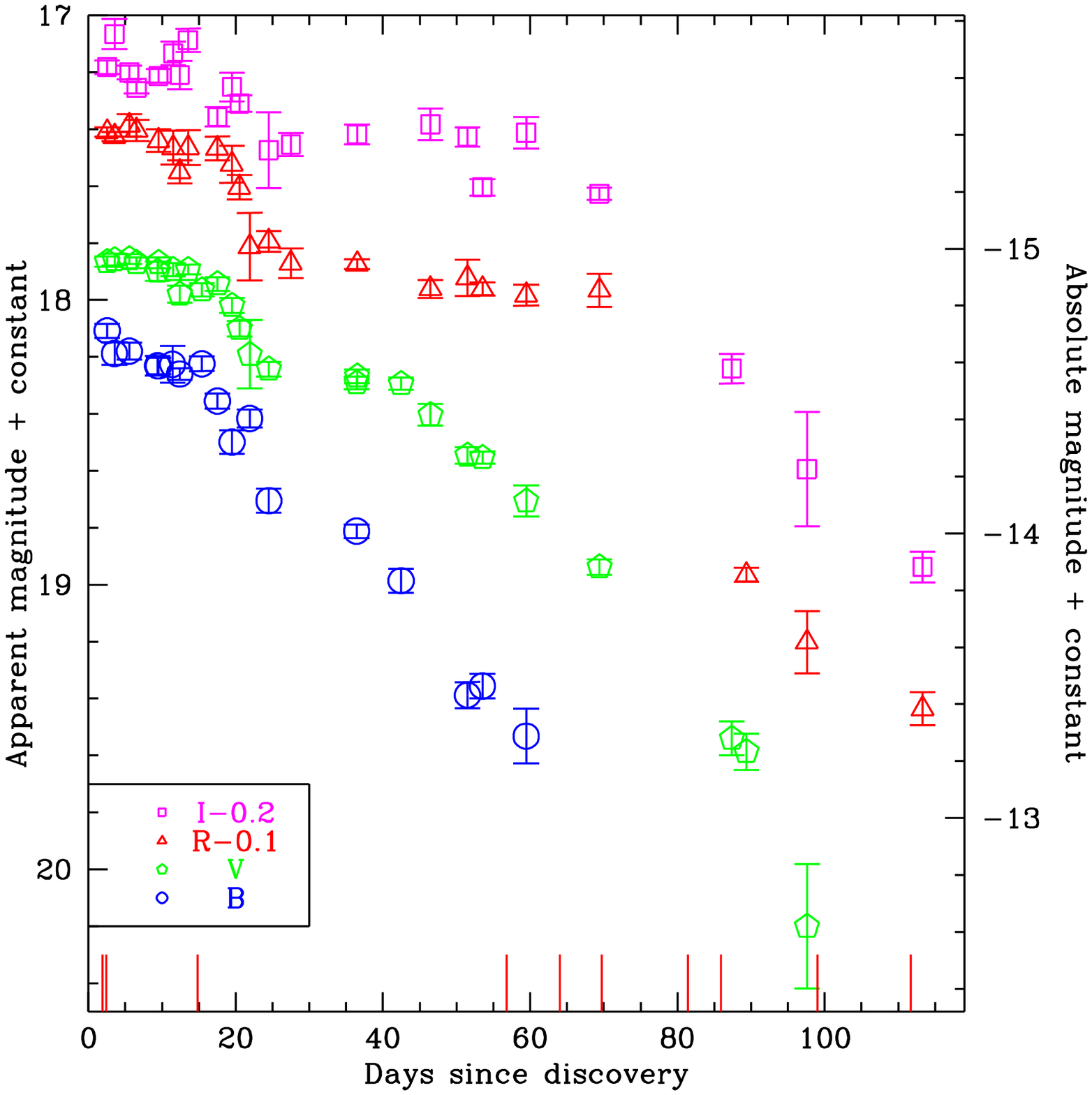}{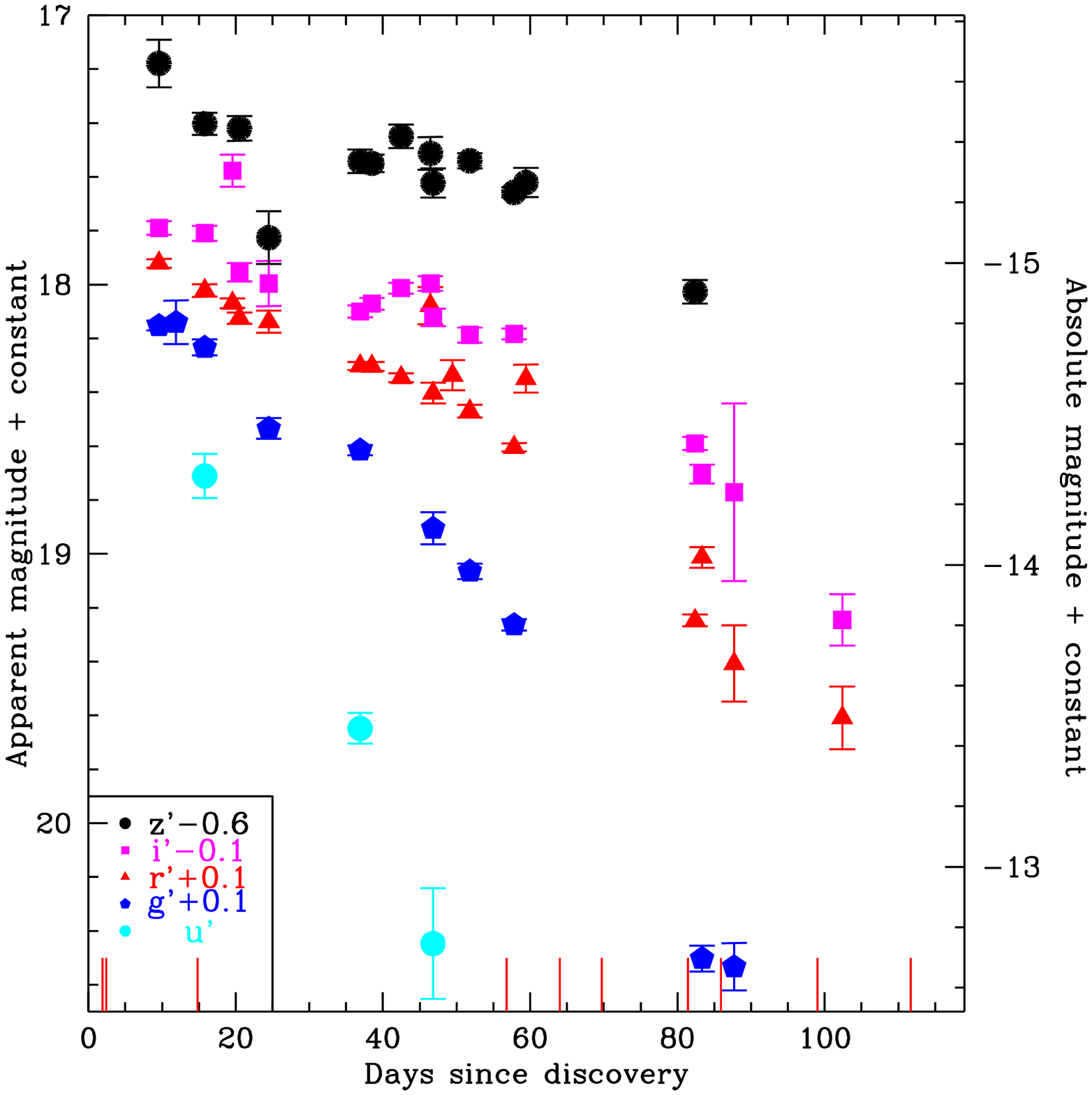}
\caption{SN~2011A $u'Bg'VRr'Ii'z'$ band light curves. The top figure represents the light curves for Landolt filters and the bottom figure for Sloan filters. Epochs are with respect to the discovery date (2011 January 2.30 UT= 0 days). Vertical red lines represent epochs of optical spectroscopy. The apparent magnitude is shown on the left hand side y--axis, while the absolute magnitude (calculated using only the distance modulus) is represented on the right hand side.}
\label{Figure.2}
\end{figure*}

\subsection{Visual--wavelength spectroscopy}

\begin{table*}
\tiny
\begin{center}
\caption{Log of optical spectroscopy of SN~2011A.\label{Table 3.}}
\begin{tabular}{l c c c c c c c c}
\tableline\tableline
JD$-2450000+$ &UT date &Epoch\tablenotemark{a} &Telescope/Instrument &Exposure(s) &Grating &Resolution FWHM (\AA) &Resolution (km s$^{-1}$) &Range (\AA)\\
\tableline
5565.8  &Jan 4.3  &2.0 &du Pont/WFCCD WF4K--1 &600 &Blue Grism  &8.0 &365 &3630--9100 \\
5566.2  &Jan 4.7  &2.4 &Gemini--North/GMOS  & 2 $\times$ 600 &B600+G5307  &5.2 & 240 &3580--9100\\
5578.7  &Jan 17.2  &14.9 &NOT/ALFOSC  &3 $\times$ 1200 &Grism 17 &1.5 &70 &6330--6900\\
5620.6  &Feb 28.1  &56.8 &SOAR/GOODMAN & 2 $\times$ 1200 &RALC 300 &14 &640 &4200--8800\\
5627.8  &Mar 7.3   &64.0 &du Pont/WFCCD   &1000 &Blue Grism &8.0 &365 &3700--9180\\
5633.8  &Mar 13.3   &70.0 &SOAR/GOODMAN  &2 $\times$ 2700 &KOSI 600 &5.1 &235 &4380--7015\\
5644.7  &Mar 24.2   &80.9 &NTT/EFOSC2  &2 $\times$ 3600 &Grism 11/16 &14.0 &640 &3740--9000\\
5649.7  &Mar 29.2   &85.9 &Baade/IMACS  &4 $\times$ 2000 &300-4.3 Grating &2.7 &120 &4190--9000\\
5662.3  &Apr 10.8   &98.5 &NTT/EFOSC2 &2 $\times$ 3600 &Grism 11/16 &14.0& 640 &3610--9050\\
5675.6  &Apr 24.1 &111.8 &SOAR/GOODMAN &2 $\times$ 2700 &KOSI 600 &7.0 &320 &4350--7015\\
5708.5  &May 27.0   &144.7 &SOAR/GOODMAN &2 $\times$ 2400 &RALC 300 &10.2 & 470 &3650--8840\\
\tableline
\end{tabular}
\tablenotetext{1}{Days after discovery 2011 January 2.30 UT.}
\end{center}
\end{table*}

Optical spectroscopy of SN~2011A was acquired using the GOODMAN spectrograph
at the Southern Astrophysical Research (SOAR) 4.1 meter Telescope, the
WFCCD at the du Pont telescope located at Las Campanas Observatory
(LCO), the GMOS spectrograph mounted on Gemini--North on the summit of Mauna Kea, EFOSC2 on the 3.6m NTT 
at La Silla, the ALFOSC imager/spectrograph on the 2.5m Nordic Optical Telescope (NOT) at La Palma and IMACS mounted on the
Baade telescope at LCO. In total we obtained 11 optical spectra mostly 
covering all the optical range, from $\sim$ 3500--9000~\AA ~except for three spectra, 
one covering only the H$\alpha$ emission and the others $\sim$ 4200--7000~\AA. 
A log of the optical observations is given in Table 3.\\

Spectroscopic reductions were performed using standard IRAF routines. All data were debiased, 
flat--fielded and cleaned of cosmic rays. Extracted 1D spectra were wavelength calibrated using 
He--Ne--Ar/Hg--Ar/Th--Ar lamps (depending on the instrument) and the calibrations were corrected 
using bright night--sky emission lines. However for the Gemini spectrum the wavelength calibration 
was computed before 1D spectrum extraction. Flux calibration was determined with spectrophotometric 
standard stars \citep{ham92,ham94}. Atmospheric absorption features were not removed except for spectra 
2.0, 64.0, and 85.9 using a telluric standard \citep{ham06}. The spectra are not corrected for Milky Way or host--galaxy extinction. The spectral sequence is shown in Fig.~3. \\

\begin{figure*}
\epsscale{0.91}
\plotone{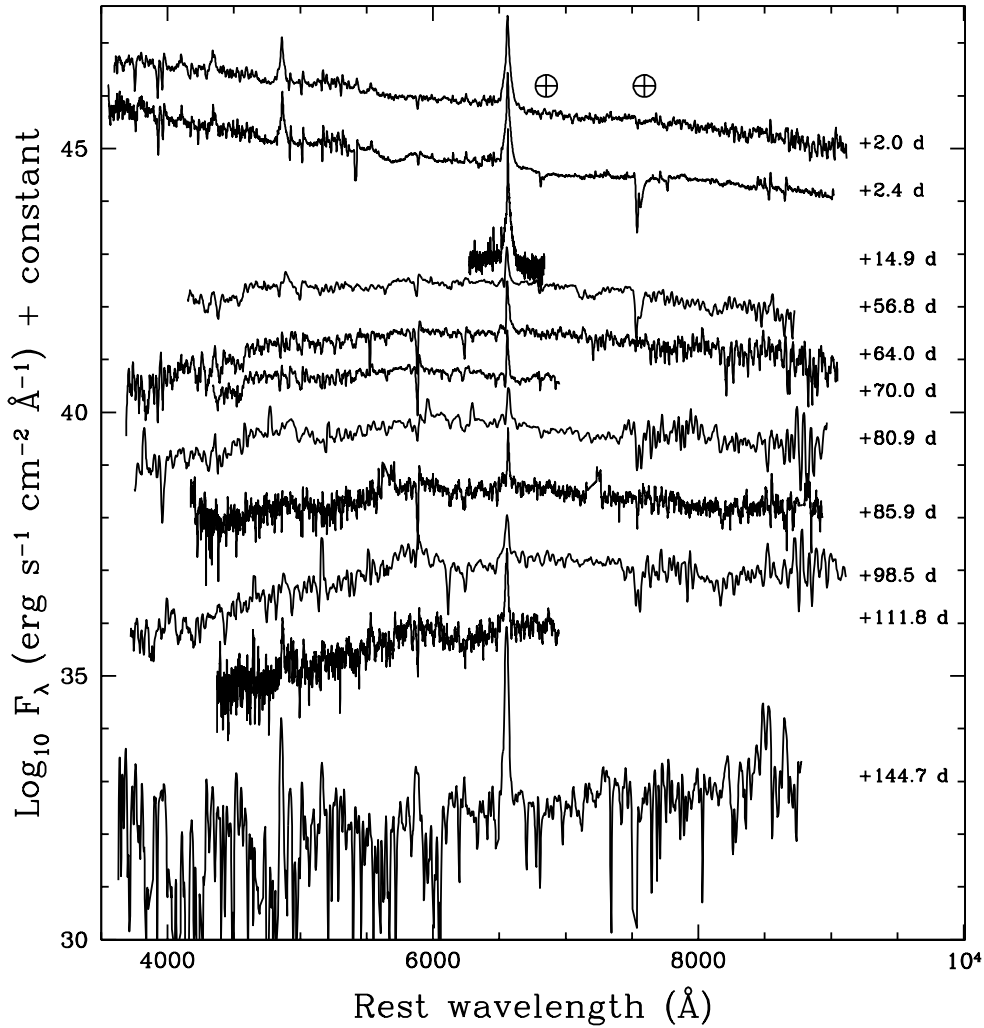}
\caption{Optical spectra of SN~2011A taken with SOAR, NTT, Gemini--North, NOT, Baade and du Pont telescopes. 
The epochs post discovery (2011 January 2.30 UT) are listed. 
Atmospheric absorption features were not removed except for spectra at 2.0, 64.0, and 85.9 days, and are indicated with a $\Earth$ symbol. Spectra were smoothed using a boxcar of 5 pixels size. Flux calibration of spectra was revised using the multi--color photometry. The spectra have not been corrected for Milky Way or host--galaxy extinction.}
\label{Figure.3}
\end{figure*}

\section{Results}

\subsection{Photometry}

\subsubsection{A rare double plateau}

One of the defining photometric features of SN~2011A, displayed in Fig.~2 is the presence of a double plateau. The $V$--band light curve exhibits an initial plateau which lasts $\sim$ 15 days with a slope $\sim$ 0.37 mag $\times$ 100~days$^{-1}$. Subsequently, the light curve declines $\sim$ 0.39 mag in $\sim$ 10 days. Then there is a second plateau phase which lasts $\sim$ 15 days with a slope of 0.29 mag $\times$ 100~days$^{-1}$, which is flatter and longer lasting at redder wavelengths. For the remaining observed epochs the light curve declines with a slope of 2.8 mag $\times$ 100~days$^{-1}$. If we look at the entire $V$--band light curve, SN~2011A declines $\sim$ 2.1 mag $\times$ 100~days$^{-1}$. 
To our knowledge this double plateau has only been seen previously in the transient SN~1997bs (see Section 4.1.2). Some other SNe such as SN~1993J \citep{richmond94}, SN~2006aj \citep{campana06}, or SN~2011dh \citep{arcavi11} present interesting characteristics (double--humped profile) which could be similar to our double plateau, but we think that the physics is most likely distinct. Indeed the double plateau seen in SN~2011A is powered by interaction whereas the double peak (in the above events) is first due to the cooling of the shock surface material and then to radioactive decay \citep{nakar14}. We note that the double plateau is not so evident in the Sloan light curves, due to less well sampled photometry and noisier data. In the Landolt filters the double plateau is clearly visible in $V$ and $R$, but less evident in $B$ and $I$. We are confident that this is a real feature of the transient.

\subsubsection{Colours}

The $(B-V)$, $(V-R)$, and $(R-I)$ colour curves of SN~2011A corrected for 
Milky Way extinction (\citealt{schlafly11}\footnote{via NED.}, $A_V$ = 0.137 mag 
and assuming $R_{V}=3.1$ \citealt{car89}) are presented in Fig.~4. 
In $(B-V)$ the object shows a nearly constant colour $\sim$0.20 mag during the
first 20 days, after which it grows steadily redder until 60 post discovery. 
A similar behavior is observed in $(V-R)$ except that the constant colour phase $\sim$0.30 mag
extends through day $\sim$40, before growing redder.
For $(R-I)$ we can see an increase from 0 to 0.4 mag in 100 days, but
the noise prevents us from ruling out an initial phase of constant color
as observed in $(B-V)$ and $(V-R)$.

\begin{figure}
\epsscale{1.20}
\plotone{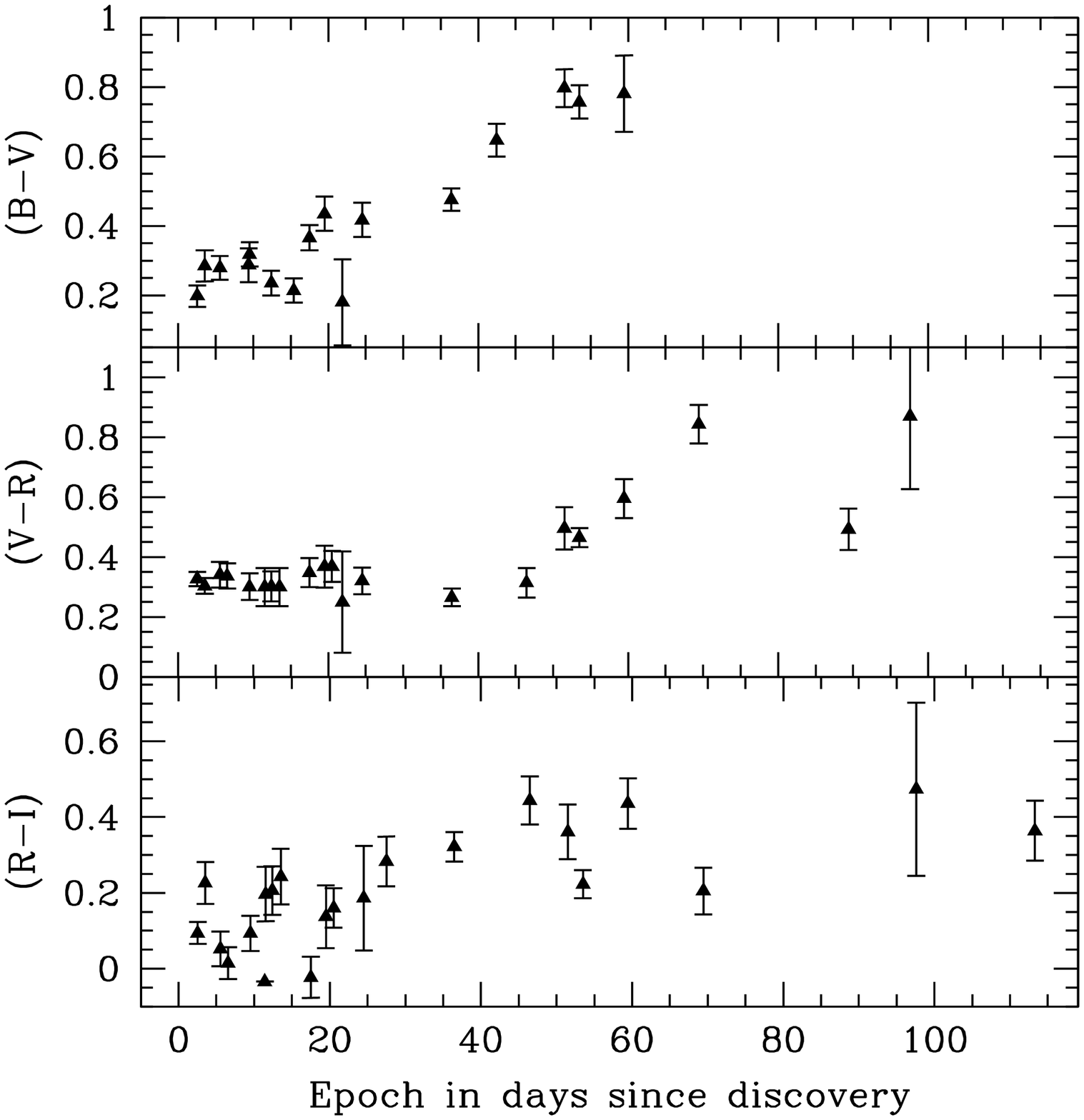}
\caption{Observed $(B-V), (V-R), (R-I)$ colour evolution of SN~2011A corrected for Milky Way extinction.}
\label{Figure.4}
\end{figure}

\subsubsection{Extinction and Absolute Magnitude}

We attempt to obtain constraints on host--galaxy extinction by analysing the early colour. 
If we assume that the initial $(B-V) = 0.2$ mag corresponds to maximum light and that the intrinsic 
colour was close to $(B-V) = 0.0$ mag, (which is what one might expect for SNe~II cf \citealt{faran14}), then 
$E(B-V)\sim 0.20 \pm 0.03$ mag (aware of the possibility that SNe~IIn may have different colours 
at maximum than normal SNe~II). Using $R_{V}=3.1$ this implies a host--galaxy extinction  
$A_{V} \sim 0.62 \pm 0.10$ mag.\\

In addition we also examined the interstellar \ion{Na}{1} D lines $\lambda\lambda 5889, 5895$ to estimate host--galaxy 
extinction using the spectrum with highest resolution taken close to discovery, i.e., that one 
obtained at Gemini--North on 2011 January 4.7 UT. The measurement of the \ion{Na}{1} D absorption 
feature equivalent width (EW) yields $EW(D_{1,2})=2.55 \pm 0.45$~\AA~. Unfortunately, 
the relation between $A_V$ and EW for unresolved \ion{Na}{1} D lines breaks down at EWs higher than 1~\AA~
\citep{phillips13}, preventing us from deriving an independent host--galaxy extinction for SN~2011A. If this detection
is due to interstellar material (ISM) then this would imply significant host--galaxy extinction. However, later we argue that the absorption is due to CSM material, and hence any connection with host galaxy--extinction is unclear.\\
\indent

Galactic extinction for SN~2011A in the $V$ band is $\sim$ 0.137 mag (\citealt{schlafly11}\footnote{via NED.}) 
assuming $R_{V}=3.1$. Using a host--galaxy redshift $z= 0.008916$ \citep{theureau07} and 
assuming a $\Lambda CDM$ cosmology with $\Omega_\Lambda=0.73$, $\Omega_M=0.27$, and 
$H_0$=73 km s$^{-1}$ Mpc$^{-1}$, we obtain a distance of 36.7 Mpc, which implies
an absolute magnitude uncorrected for host extinction of $M_{V}=-15.10$ mag close to discovery. Aware of the uncertainties in the $A_{V}$ from the host--galaxy ($0.62 \pm 0.10$ mag), we obtain an absolute $V$--band magnitude of $-15.72$ mag. 
This is low for the SN~IIn class ($-16\geq$ $M_{V}$ $\geq-20$, \citealt{richardson02}), and is high compared to the SN~impostors (e.g., SN~1997bs has $M_{V}>-14$ \citealt{van00} or see compilation in \citealt{smith11}). 
It is therefore difficult to differentiate between a SN~IIn or SN~impostor origin 
for this transient using solely its $M_{V}$.\\

\subsection{Spectroscopy}

As expected for an interacting transient, the spectral sequence presented in Fig.~3 
is dominated by relatively narrow H$\alpha$ emission. However, there is significant 
evolution with time, as will now be discussed. 

\subsubsection{Spectral evolution}

The spectrum with the highest resolution taken close to discovery (+2.4 days), is 
dominated by Balmer lines, most prominently H$\alpha$ and H$\beta$ in emission,
with a characteristic width of $\sim$2400 km s$^{-1}$ measured using the Full Width at Half Maximum (FWHM). In what follows we will refer to these lines as ``broad'', although the reader should keep in mind that
these velocities are much lower than what is observed in most SNe II ($v$ $\sim10{^4}$ km s$^{-1}$). 
We observe also many narrow ($\sim$ 500--1000 km s$^{-1}$) P--Cygni 
profiles attributed to \ion{Ca}{2} $\lambda\lambda 3933,3968$, the \ion{Ca}{2} near--infrared triplet 
$\lambda\lambda 8498, 8542, 8662$, and many \ion{Fe}{2} lines.
After $\sim$ day 50, H$\alpha$ emission is still present but H$\beta$ 
is significantly weaker. We see a similar evolution of the \ion{Fe}{2} lines which 
are present during the first few days, but disappear from day 56.8 onwards.
The \ion{Ca}{2} triplet is not clearly seen in spectra between days 56.8 and 85.9 
due to noisy data but is still visible on the +98.5 day spectrum, and possibly
144.7 days after discovery. \\
\indent 

To understand SNe~IIn, it is common practice to analyse the H$\alpha$ line profile 
and its evolution. This is shown in Fig.~5 where we see considerable temporal variations.
Indeed, from discovery to 56.8 days later, our spectra are characterised by prominent 
``broad'' emission ($v$ $\sim2000$ km s$^{-1}$). Then between 56.8 days and 85.9 days post discovery, a narrow low--velocity (600--1100 km s$^{-1}$) P--Cygni absorption appears. After day 85.9 we see again a ``broad'' 
emission with no signs of the P--Cygni profile. Again, we emphasize the absence in the spectra 
at all epochs of any kind of high velocity components ($\sim 10{^4}$ km s$^{-1}$)
characteristic of SN ejecta.\\

\begin{figure}
\epsscale{1.20}
\plotone{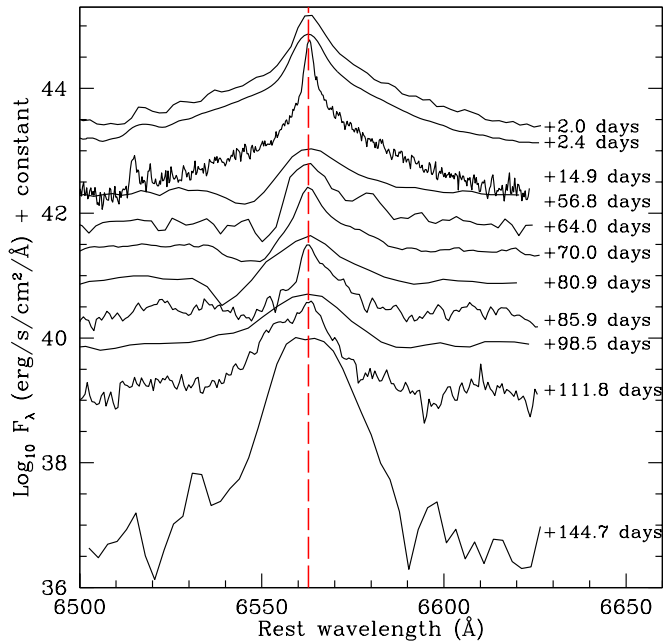}
\caption {H$\alpha$ line profile evolution of SN~2011A. The epochs are indicated in black and are with respect to the discovery date (2011 Jan 2.30 UT = 0 days). The dotted red vertical line represents the rest--frame H$\alpha$ wavelength, at 6562.81~\AA.}
\label{Figure.5}
\end{figure}

In Fig.~6 we show the H$\alpha$ line evolution splitting it into epochs with and 
without P--Cygni profiles. The P--Cygni component is seen in spectra obtained 
on days +56.8, +64.0, +70.0, and +80.9. The blueshifted absorption velocity  
has values between 575 km s$^{-1}$ to 1060 km s$^{-1}$.
It is unlikely that these low--velocity P--Cygni profiles can be 
attributed to the ejecta, which is expected to have velocities of several 
thousand km s$^{-1}$. Therefore, we argue that this P--Cygni component is related 
to a pre-SN wind. The wind velocity is consistent with an LBV progenitor 
(100 -- 1000 km s$^{-1}$ \citealt{smi07a}); but too high for red supergiants 
(20 -- 40 km s$^{-1}$; \citealt{smi07b}) and possibly too low for Wolf--Rayet stars 
(1000 -- 5000 km s$^{-1}$; \citealt{abb87}).

\begin{figure}
\epsscale{2.40}
\plottwo{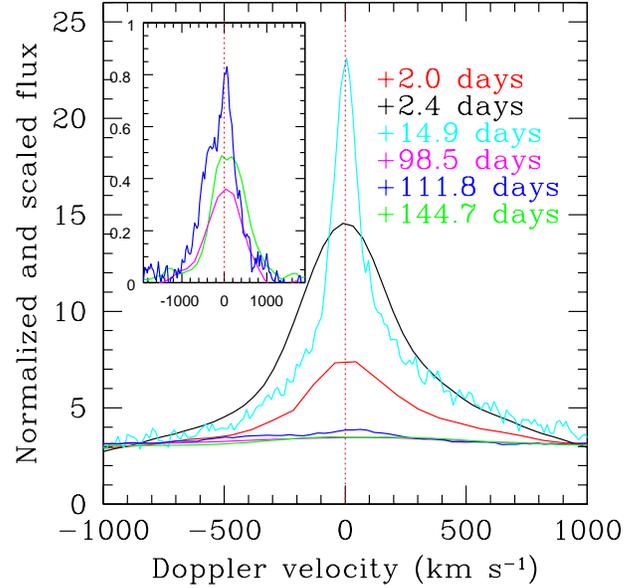}{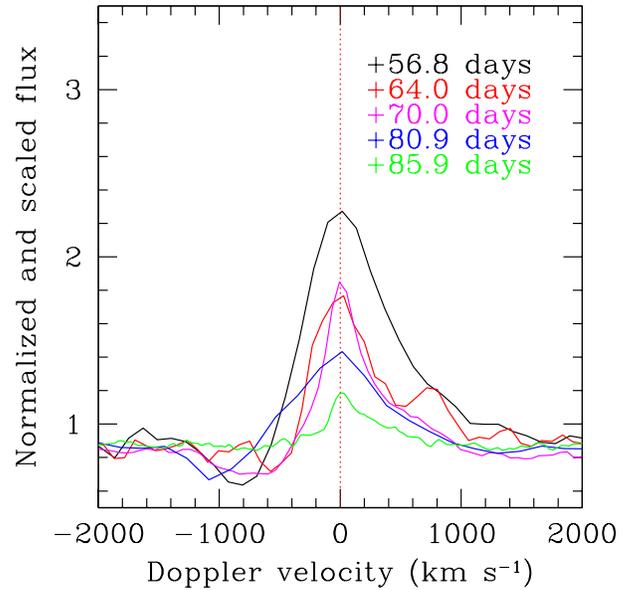}
\caption{H$\alpha$ line profile evolution. The top figure shows the
 H$\alpha$ line profile showing a broad component without absorption. We also present a zoom on H$\alpha$ for late spectra. The bottom figure shows the H$\alpha$ line profile with P--Cygni absorption. We normalized all spectra with respect to the Gemini spectrum continuum. The scale has been adjusted to overlap the broad wings of H$\alpha$.}
\label{Figure.6}
\end{figure}
\indent

Our first spectra, taken 2.0 days and 2.4 days after discovery, show many strong \ion{Fe}{2} 
features with P--Cygni profiles. The velocity measured with respect to the emission 
from the \ion{Fe}{2} multiplet $\lambda\lambda 4923, 5018, 5169$ lines in the +2.4 day spectrum 
is between 430-460 km s$^{-1}$, arguing for a CSM rather than ejecta origin. 
We also note that these velocities have the same order of magnitude as those derived 
from the H$\alpha$ P--Cygni component. The same is seen in the red part with the calcium 
triplet $\lambda\lambda 8498, 8542, 8662$. Indeed in the +2.4 day spectrum we measured 
low velocity with respect to the emission ($\sim$ $440$ km s$^{-1}$), again supporting 
the hypothesis of CSM origin and consistent with the velocities derived from the \ion{Fe}{2} lines. Looking at the continuum we can also estimate the temperature through a blackbody fit. The temperature evolution shows a rapid decline from $\approx$ 9000 K in the +2.0 days spectrum to $\approx$ 5000 K in the +56.8 days spectrum and to $\approx$ 3300 K at the later epochs. 

\subsubsection{Variable low velocity absorption close to \ion{Na}{1} D}

As seen in the Type IIn SN~1994W \citep{chu04}, we observe a strong absorption near the \ion{Na}{1} D doublet wavelength 
($\lambda\lambda 5889, 5895$) which becomes stronger with time and has low velocity. This is shown in Fig.~7, 
where the evolution of this feature is presented. The equivalent width and velocity of the line profile were calculated 
for each epoch. Because the majority of our spectra were taken in low resolution, the sodium doublet is unresolved, hence, we will refer to this feature as absorption at the wavelength of \ion{Na}{1} D, 
due to the fact that the \ion{Na}{1} D can be blended with the \ion{He}{1} $\lambda 5876$ line (we also note the possible presence of \ion{He}{1} at the wavelength $\lambda 6678$ and $\lambda 7065$ in the 2.4 days spectrum). Therefore the total blended doublet is measured as a single line. In Fig.~8 we present both the equivalent width and velocity evolution and see that the absorption becomes stronger with time. Indeed the equivalent width is initially equal to $\sim$ 3~\AA~ and increases to 10~\AA~ after the first 70.0 days. 
At the same time, the feature profile evolves. In the first two spectra, only an absorption line is observed while, between days 56.8 and 110.3 a P--Cygni profile appears. The velocities are measured from the \ion{Na}{1} D center, 5892~\AA~ with respect to the emission are shown in Table 4. Fig.~8 also shows the average \ion{Na}{1} D equivalent widths from ISM for all the \textit{Carnegie Supernova Project} (CSP; \citealt{ham06}) SNe~II (as used in \citealt{and14}). We measure the equivalent width for each SN regardless of the epoch and we averaged these values.
This is again further evidence of a CSM origin for this line: the strength is much higher than seen in any other SNe~II 
from the CSP database. Even if the low resolution spectra prevent us to be sure that the measured absorption close 
to \ion{Na}{1} D position is only due to \ion{Na}{1} D and not blending with other lines, the absorption strength is unprecedented.
 
\begin{figure}
\epsscale{1.20}
\plotone{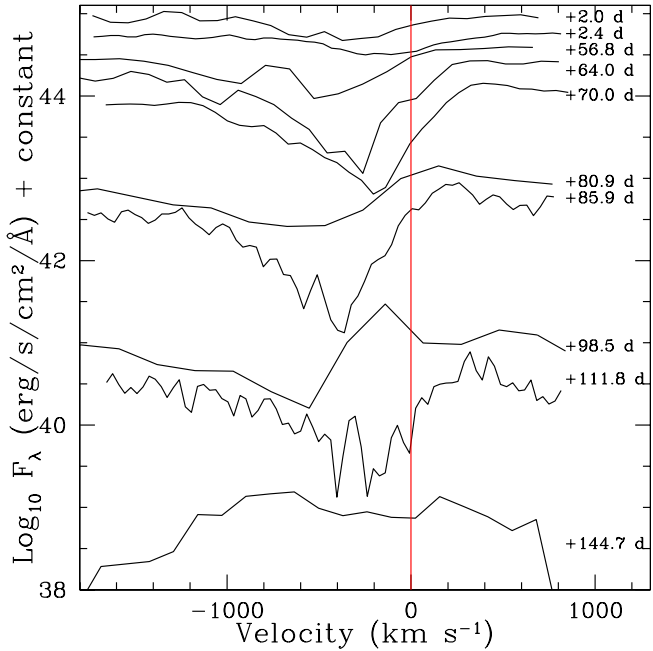}
\caption{Zoom on the absorption close \ion{Na}{1} D  of our SN~2011A optical spectra. The epochs are listed with respect to the discovery date (2011 January 2.30 UT= 0 days). The red line represents the center of the \ion{Na}{1} D corrected to the redshift of the host--galaxy. We note that this appears offset by $\sim$ 200 km s$^{-1}$ from the emission component of \ion{Na}{1} D.}
\label{Figure.7}
\end{figure}

\begin{table}
\tiny
\begin{center}
\caption{Evolution of the absorption close \ion{Na}{1} D with time.\label{Table 4.}}
\begin{tabular}{lccc}
\tableline\tableline
Epoch\tablenotemark{a} &Equivalent width (\AA)\tablenotemark{b} &Velocity (km s$^{-1}$)\tablenotemark{b} &Line profile\\
\tableline
 & & &\\
2.0   &3.2(0.5)  &530(25) &absorption\\
2.4   &2.6(0.7)  &400(30) &absorption \\
56.8  &6.9(0.35)  &500(75) &P--Cygni \\
64.0    &11.5(0.7)  &510(55) &P--Cygni\\
70.0  &9.5(0.3)  &490(70) &P--Cygni \\
80.9  &8.0(1.1)  &820(50) &P--Cygni\\
85.9  &10.8(0.5) &600(75) &P--Cygni\\
98.5   &7.7(0.8)  &830(125) &P--Cygni\\
111.8 &10.6(1.6) &400(130) &P--Cygni\\
144.7 &$\cdots$          &$\cdots$       &$\cdots$ \\

\tableline
\end{tabular}
\tablenotetext{1}{Days after discovery 2011 January 2.30 UT.}
\tablenotetext{2}{We note in parenthesis the associated error. The errors come from the uncertainties of defining the continuum for measurement of the absorption and emission components. This was achieved using multiple measurements changing the continuum position each time. The velocities are measured from the \ion{Na}{1} D center, 5892~\AA~with respect to the emission.}

\end{center}
\end{table}

\begin{figure}
\epsscale{1.2}
\plotone{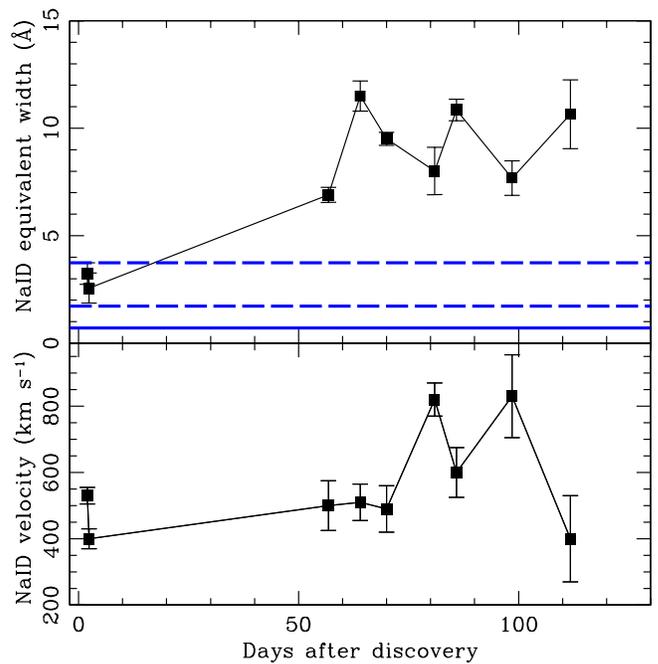}
\caption{Top figure: the absorption close \ion{Na}{1} D equivalent width evolution with time for SN~2011A. The solid blue line shows 
the average \ion{Na}{1} D EWs (averaged over time) for all CSP SNe~II (including SNe~IIn, but not SNe~IIb) where the EW is most likely of ISM origin. 
The lower dashed blue line shows the 1 sigma error on the EW estimation, and the upper shows 3 sigma error. The lower panel shows 
the absorption close \ion{Na}{1} D velocity evolution measured with respect to the emission.}
\label{Figure.8}
\end{figure}

\subsubsection{Wind velocity}

In order to measure the wind velocity we estimate the width of the narrow component of the H$\alpha$ emission line.
For this estimation we decompose the H$\alpha$ feature using a least-squares PYTHON script, which solves the best--fit multi Gaussian decomposition. The best fit was using three components, so nine parameters
 are used as input: the FWHM, the amplitude and the center of each Gaussian corresponding to the three components. 
Using these values our script provides the best--fit Gaussian parameters, the best FWHM, the amplitude and the center of each component. The errors on these parameters are derived assuming a reduced chi squared equal to one. In Fig.~9 we show the H$\alpha$ emission line decomposition for spectra taken with ALFOSC (highest resolution spectrum available) on 2011 January 17.2 UT. Our three components fit the H$\alpha$ emission line yields velocities of 107 $\pm$ 2 km s$^{-1}$, 467 $\pm$ 14 km s$^{-1}$, and 1971 $\pm$ 35 km s$^{-1}$. Each velocity component is low compared to other SNe~IIn. For comparison, \citet{kie12} found typical velocities of 2000 km s$^{-1}$ for the intermediate component. We also note that \citet{pas04} found that for low--luminosity SNe~II the expansion velocities range from 3000--5000 km s$^{-1}$ during early epochs. Therefore, SN~2011A appears to have ejecta velocities lower than the majority of normal SNe events. This could be evidence for a SN~ impostor origin, as will be discussed below. In addition, recently \citet{tartaglia14} have reported a case of an interesting transient, SN~2007sv. From this object the authors, also analysing the H$\alpha$ emission line, derived very similar velocity values. Indeed the broad component velocity was estimated to 2000 km s$^{-1}$ (1971 km s$^{-1}$ for SN~2011A), an intermediate component velocity of 600--800 km s$^{-1}$ (467 km s$^{-1}$ for SN~2011A) and a narrow component velocity of 120--150 km s$^{-1}$ (107 km s$^{-1}$ for SN~2011A). In their paper, they conclude that the transient SN~2007sv is most likely a SN~impostor, based mainly on its absolute magnitude. Note that the absolute magnitude derived for this transient is $\sim$~2 mag fainter than SN~2011A.\\
\indent

Note that from the Gaussian decomposition of the highest resolution spectrum we can calculate the progenitor mass--loss rate using the following formula \citep{chu94,kie12}:

\begin{equation}
L_{H_{\alpha}}=\frac{1}{2} \dot{M} \psi \frac{v_{shock}^{3}}{v_{wind}}
\end{equation}

where $\psi$ is the efficiency of the conversion of mechanical energy into optical energy in the shock wave 
(we adopt as \citealt{kie12}, $\psi$= 0.1). $\dot{M}$ is the mass loss rate, $v_{wind}$ is the unshocked wind velocity derived as described above and $v_{shock}$ is the shock velocity. The shock velocity was taken as the FHWM of the intermediate width component. The absolute $L_{H_{alpha}}$ was then obtained by integrating over the intermediate width feature and taking account of the distance to the SN. We used the same luminosity distance used above, i.e, 36.70 Mpc.
We found $\dot{M}$ = 0.038 M$_{\sun}$ yr$^{-1}$ with a wind velocity of 110 km s$^{-1}$ and a shock velocity about 470 km s$^{-1}$. This mass loss rate is consistent with values found by \citet{kie12} and is too high for any class of massive stars other than LBVs in the eruptive phase \citep{hum94}.

\begin{figure}
\epsscale{1.3}
\plotone{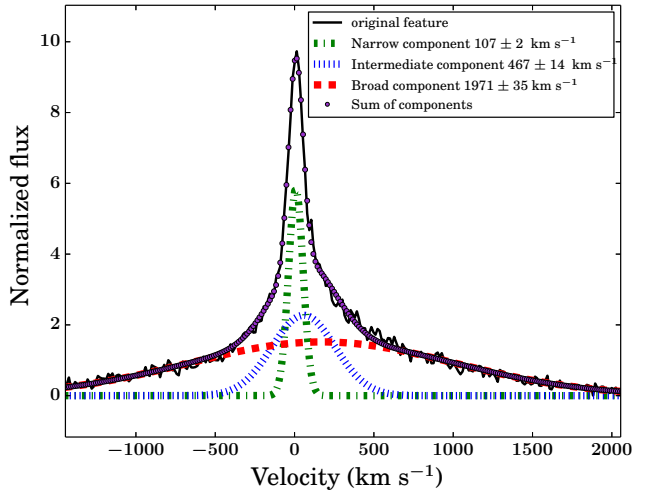}
\caption{The decomposition of the H$\alpha$ emission line for the spectrum taken with ALFOSC at NOT on 2011 January 17.2 UT.}
\label{Figure.9}
\end{figure}

\section{Discussion}

SN~2011A shows a number of interesting properties. In order to understand the nature of the transient, 
we compare in this section our object with other interacting transients. We discuss the transient in comparison to other events which possibly show similarities. However, in any case where similarities exist, we show that there are other features which confirm the uniqueness of SN~2011A.\\ 

\subsection{Photometry}

\subsubsection{Photometric comparison to other SNe~IIn}
As stated in the introduction, the SN~IIn group is very heterogeneous, in terms of luminosity, light curve shape and spectral evolution. Based on \citet{kie12} who proposed to differentiate SNe~IIn according to their photometric and spectroscopic characteristics, \citet{kan12} 
identify a subclass of SNe~IIn in which they included SN~2009kn \citep{kan12}, SN~1994W \citep{chu04} and SN 2005cl \citep{kie12}. To see if SN~2011A shares common properties with these SNe and belongs to this subclass, we compare the light curves and spectra (see Section 4.2.2). We also include comparison with SN~1998S \citep{fas01} a well studied SN~IIn, SN~2005kj \citep{tad13} which, similarly to SN~2011A shows evidence of low velocity absorption at \ion{Na}{1} D wavelengths, SN~1997bs \citep{van00} classified as SN~impostor (note that we compare in more detail SN~2011A and SN~1997bs in the Section 4.1.2.), and SN~2007sv, also recently classified as an SN~impostor \citep{tartaglia14}. These comparisons are shown in Fig.~10, where the $V$--band light curves are plotted. Note that we do not present the SN~1994W light curve due to the lack of data and because it is very similar to SN~2009kn \citep{kan12}. 
As we can see in Fig.~10 SN~2009kn is brighter than SN~2011A with $M_{V}$ $\simeq$ $-$17.5/$-$18 and the SN~2009kn light curve does not show a double plateau like SN~2011A. Indeed, after an initially declining plateau phase with a slope equal to 0.018 mag day$^{-1}$, we see a quick fall around 0.08 mag day$^{-1}$ followed by a slow decline phase with a slope of $\sim$ 1 mag 100~days$^{-1}$. However, the photometry of SN~2009kn does not cover the first few tens of days from the assumed explosion date. As can be seen in Fig.~10, SN~2011A is photometrically very different that the other SNe~IIn. None of the other selected SNe~IIn show a double plateau and they are all $\geq$ 2 mag brighter than SN~2011A. Also we remark that SN~2011A is $\sim$ 2 mag brighter than the SN~impostors. Fig.~10 allows us to conclude that SN~2011A is a very rare event which does not share any similarities in the light curve with other SN~1994W--like SNe~IIn. An important additional point to note is the origin of the two SNe~IIn; SN~1994W and SN~2009kn used to compare SN~2011A to other SNe~IIn. Indeed the origin is not so clear, for example \citet{dessart09} argued that maybe SN~1994W was not a SN, but rather the result of two interacting shells of material related to LBV--like eruptions. Also, in the case of SN~2009kn \citep{kan12} a non--SN origin could not be completely excluded, but also found the observations to be consistent with an electron capture SN. We note that later an argument will be presented that constrains the epoch of the photometry
of SN~2011A to be $\sim$ 50 days later than shown in Fig.~10. However, the main conclusions remain valid: SN~2011A 
has a luminosity between the two classes of interacting transients, SNe~IIn and SN impostors; this event shows a rare
double plateau.
\begin{figure}
\epsscale{1.2}
\plotone{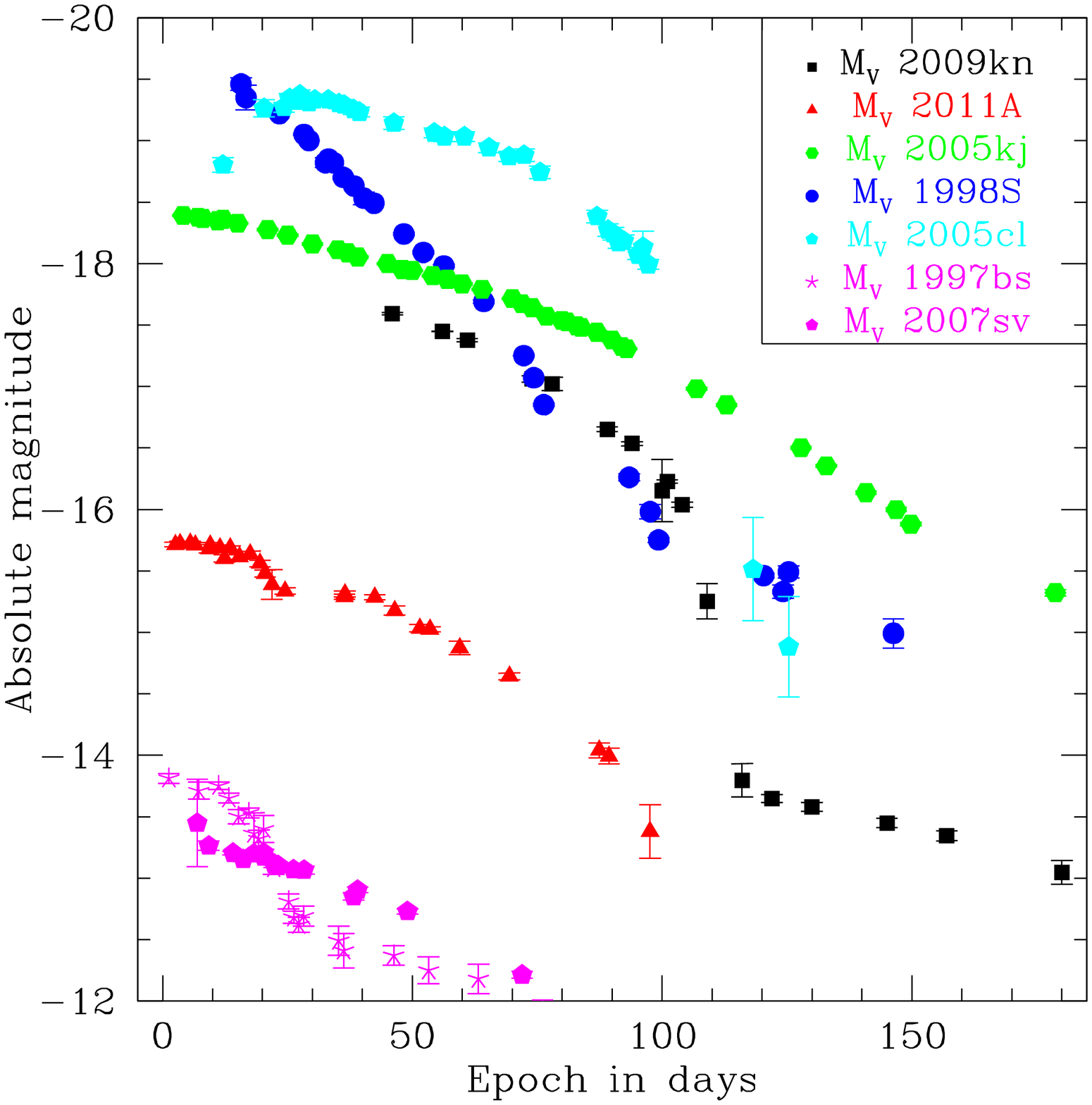}
\caption{Comparison of the absolute $V$--band light curves of SN~2011A, SN~1998S \citep{fas01}, SN~2009kn \citep{kan12}, SN~2005cl \citep{kie12}, SN~2005kj \citep{tad13}, SN~1997bs \citep{van00}, and SN~2007sv \citep{tartaglia14}. Each SN was corrected for Milky Way extinction. Full red triangles represent SN~2011A corrected for host--galaxy extinction ($A_v = 0.62$ mag). For SN~2009kn we use the explosion date estimation from Kankare et al. (2012). For SN~1998S, SN~2005kj, SN 2005cl, SN~1997bs, and SN~2007sv we use the discovery date as the explosion date. For each SN we applied the host--galaxy extinction estimated by the authors.}
\label{Figure.10}
\end{figure}
\\
\indent

In Fig.~11 we present the optical colours of SN~2011A together with the colour evolution of SN~2009kn, one SN~1994W--like, and the well-studied SN~1998S. We do not present the SN~1994W and SN 2005cl colours due to the lack of data. All magnitudes were corrected for Milky Way extinction. Taking the discovery date as the explosion date, the colour evolution of SN~2011A looks fairly similar to SN~1998S. Indeed, the $(B-V)$ colour becomes redder in SN~1998S between days 0 and 60. However, in $(V-R)$, SN~1998S does not show an initial plateau as SN~2011A.\\
\indent

The colour evolution of SN~2009kn looks similar to SN~2011A in $(B-V)$ and $(V-R)$, but shifted by $\sim$ 50 days. This could be explained by the lack of precision in the explosion date for SN~2011A. If we shift our epochs by 50 days (motivated by the spectral analysis in section 4.2.1), the colour evolution in $(B-V)$ and $(V-R)$ is very similar to SN~2009kn. Indeed for $(V-R)$, SN~2009kn shows a small plateau during $\sim$ 50 days and then an increase from 0.3 to 0.7 mag. It is more difficult to compare the behavior of the $(R-I)$ colour, due to their larger errors, however we can see the same trend: for SN~2011A, $(R-I)$ increases from $\sim 0$ to $\sim 0.4$ mag in 100 days, and for SN~2009kn, $(R-I)$ increases from $\sim 0.2$ to $\sim 0.55$ mag in the same temporal windows. The colours between SN~2011A and SN~2009kn are very similar, but shifted by 50 days which could be consistent with the hypothesis proposed in section 4.2.1, namely, that the explosion date is not close to the discovery date but more consistent with 50 days earlier.

\begin{figure}
\epsscale{1.20}
\plotone{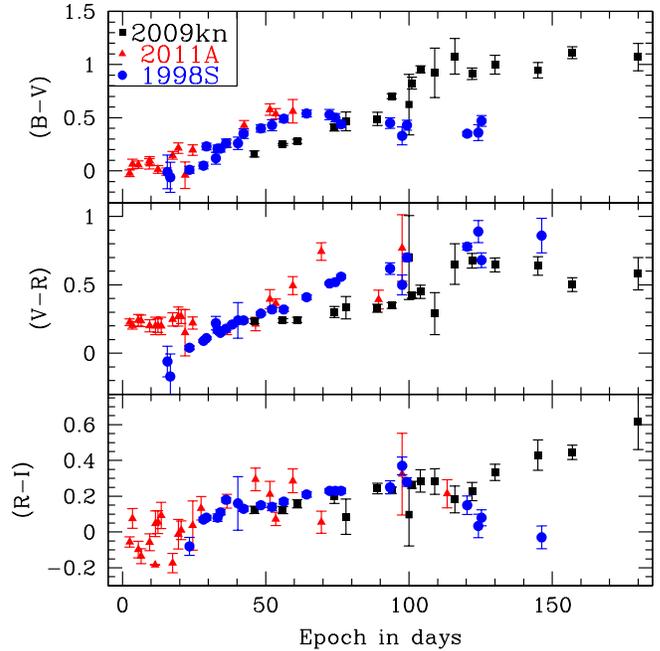}
\caption{$(B-V)$, $(V-R)$, $(R-I)$ colour evolution of SN~2011A (red triangles) compared to other SNe, SN~2009kn (Kankare et al. 2012) shown in black squares and SN~1998S (Fassia et al. 2000) represented by blue circles. All colours were corrected for Milky Way extinction and by host--galaxy extinction. For SN~2011A and SN~1998S the epochs are with respect to the discovery date, whereas for SN~2009kn they are with respect to the explosion date estimated by the authors.}
\label{Figure.11}
\end{figure}

\subsubsection{SN~2011A versus SN~1997bs}

The light curve comparison between SN~2011A  and SN~1997bs (which was classified as a SN~impostor by \citealt{van00}) shows similarities. The light curves of both objects in $B$, $V$, $R$, and $I$ bands are shown in Fig.~12. For the SN~1997bs explosion date, we choose 1997 April 15th (UT discovery date), which should be valid because an image taken on 1997 April 10th does not show anything at the location of the SN. During the first 15 days of evolution in $R$ and $I$ bands both objects exhibit a plateau. After this first plateau the two transients show a second plateau but the drop between the two is much larger for SN~1997bs: $\sim$ 0.5 mag for SN~2011A and $\sim$ 1 mag for SN~1997bs in $R$. The $B$ and $V$ bands have also some similarities. Note, while the light curve evolution displayed in Fig.~12 appears very similar, this may be coincidental given the uncertainty of the SN~2011A explosion date. However, the important observation is the similar double plateau features in both light curves, which may be evidence for the same underlying mechanism; namely interaction with two separate CSM shells. 
The optical colours of SN~2011A compared with SN~1997bs are presented in Fig.~13. From the $(B-V)$ colour we can compare the colours near maximum. For SN~2011A we find $\sim$ 0.25 mag whereas for SN~1997bs \citet{van00} found 0.67 mag, adding a host extinction for 1997bs. Additionally we can compare the colour evolution. In the $(V-R)$ colour, the evolution looks similar during the first 30 days. For SN~1997bs, a plateau at 0.3 mag during 30 days is seen and then a reddening as for SN~2011A. But in SN~2011A the plateau lasts 50 days. Despite the noise, $(B-V)$ colour exhibits the same trend with a plateau during 20 days and then a reddening. Again, in $(R-I)$, SN~2011A and the SN~impostor show the same evolution with a reddening during the first 40 days. Although the data are noisy the shape of the reddening is pretty similar. Finally, note that the SN~impostor nature of SN~1997bs is not entirely clear as there is some evidence that the progenitor did not survive to the explosion \citep{li2002,adams15}. Hence, while SN~2011A shows similarities to SN~1997bs, this is not concluding proof that the former was an imposter event.

\begin{figure*}
\epsscale{0.9}
\plotone{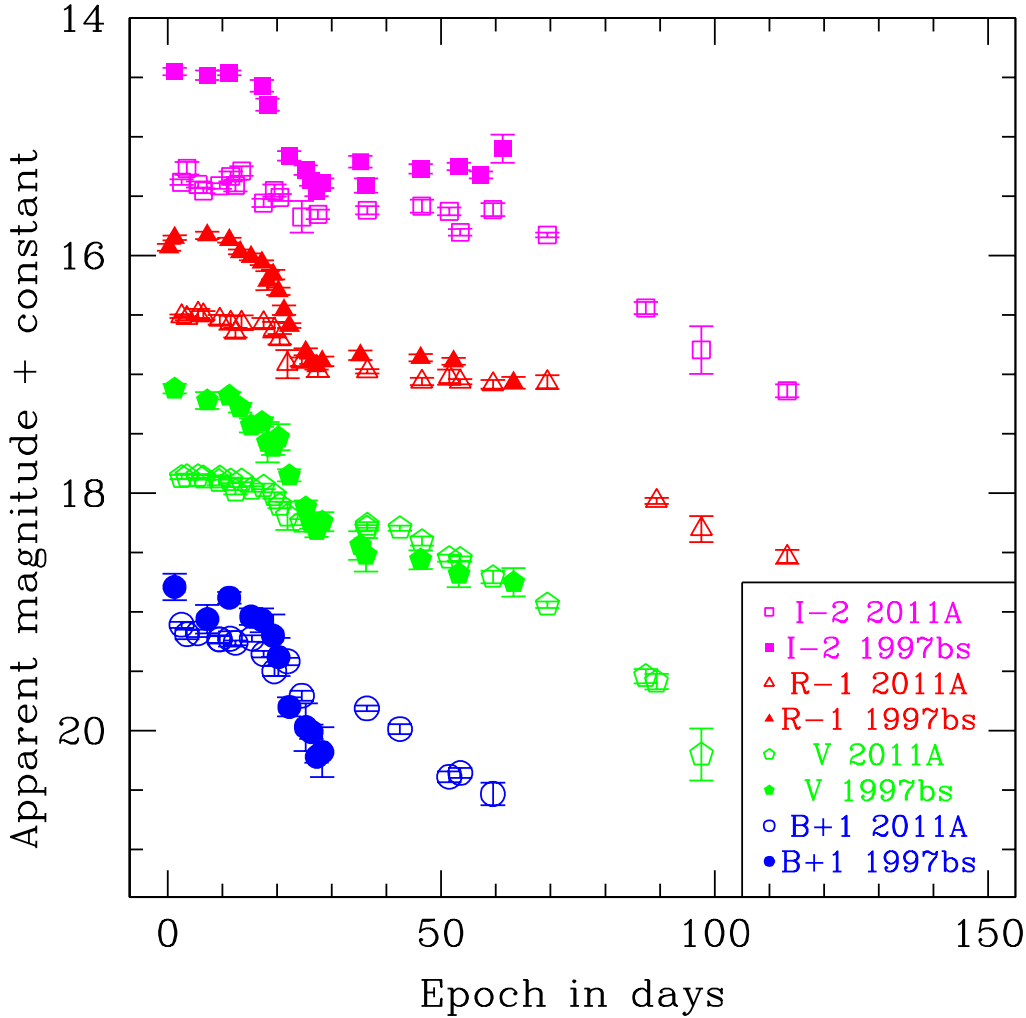}
\caption{Comparison of the $BVRI$ light curves of SN~2011A and the impostor SN~1997bs. Photometric data of the SN~impostor were taken from \citet{van00}. For each band empty symbols represent SN~2011A and full symbols SN~1997bs. Squares represent $I$-band light curve, triangles $R$ band, hexagons $V$ band and circles $B$ band.}
\label{Figure.12}
\end{figure*}

\begin{figure}
\epsscale{1.20}
\plotone{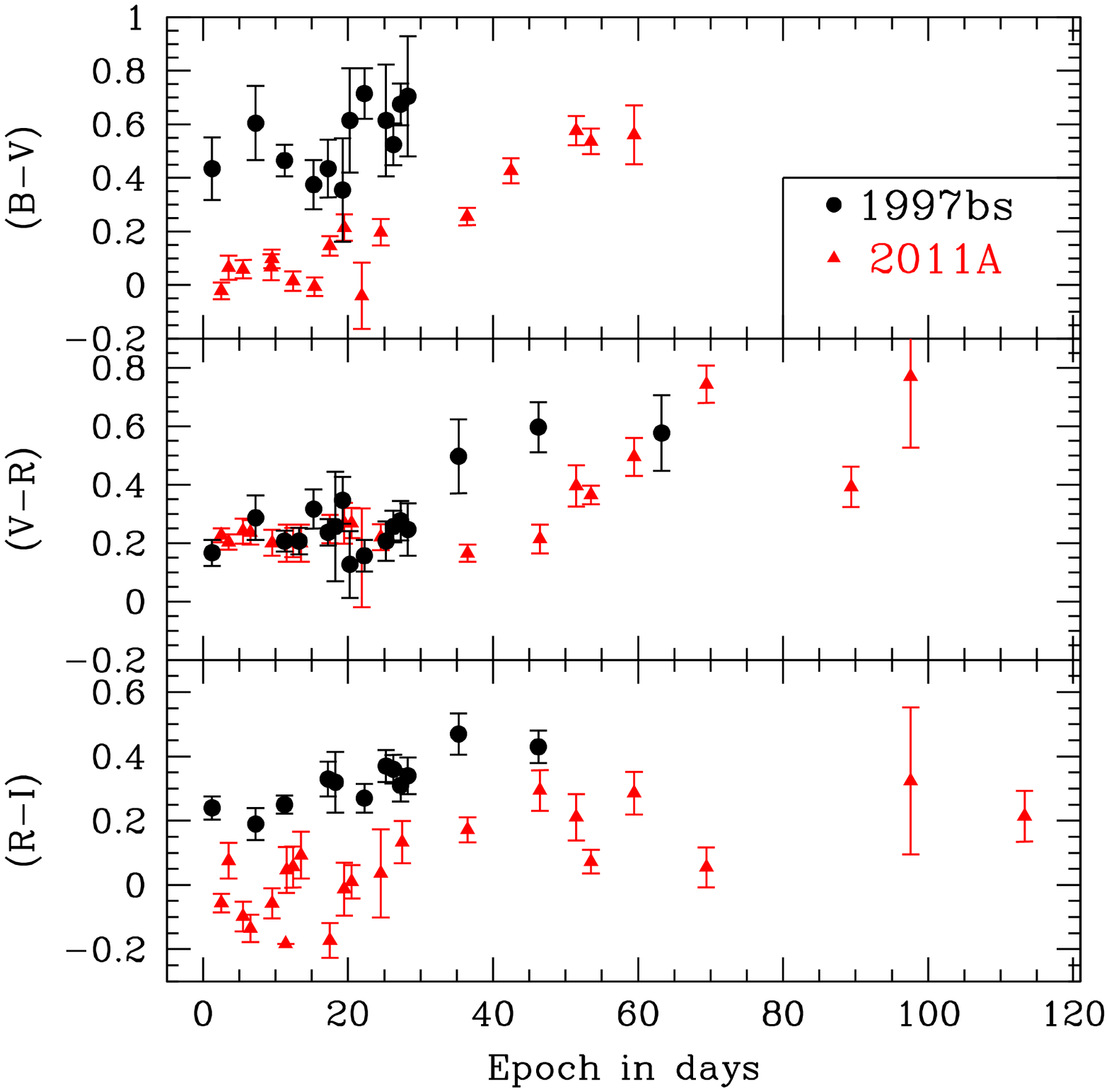}
\caption{$(B-V)$, $(V-R)$, $(R-I)$ colour evolution of SN~2011A (red triangles) compared to impostor SN~1997bs \citep{van00} shown in black circles. All colours were corrected by Milky Way extinction and by host--galaxy extinction.}
\label{Figure.13}
\end{figure}

\subsubsection{$^{56}$Ni mass estimate}

One diagnostic to discriminate between SNe and SN~impostors is the amount of synthesized $^{56}\mathrm{Ni}$. For low mass core--collapse SNe (SNe~IIP) one expects to find a $^{56}$Ni mass around 0.0016 -- 0.26 ${\rm M}_{\odot}$ \citep{ham03a} whereas for non--terminal explosions we do not except any $^{56}$Ni production because only the outer layers of the star are ejected. For very massive CCSNe produced by progenitors with initial masses between 30 -- 100 ${\rm M}_{\odot}$, $^{56}$Ni masses between 2.3 -- 6.6 ${\rm M}_{\odot}$ have been predicted \citep{ume08}, and masses between 0.07 -- 0.6 ${\rm M}_{\odot}$ have been estimated for progenitor masses between 16 -- 46 ${\rm M}_{\odot}$ \citep{mazzali09}.
We can obtain a $^{56}$Ni mass estimate using the formula from \citet{ham03a} and the last $V$--band magnitude (epoch 97.6 days). This value should be taken as an upper limit of the $^{56}$Ni mass because the spectrum at this epoch does not indicate that the emission is being solely powered by radioactivity. Indeed in this spectrum narrow lines are still present due to interaction.\\
\indent 

As stated previously the explosion date is not well constrained and could be 50 days before discovery. Therefore the $^{56}$Ni mass is calculated assuming the explosion date as the discovery date and also using an explosion date 50 days before discovery. We find 0.01 -- 0.015 ${\rm M}_{\odot}$ without extinction and 0.03 -- 0.05 ${\rm M}_{\odot}$ with extinction. In both cases the upper value is using an explosion date 50 days before discovery. These values are consistent with \citet{ham03a} and lower than \citet{mazzali09}. Indeed Hamuy (2003) found for SNe~II values between 0.0016 -- 0.26 ${\rm M}_{\odot}$ and Mazzali et al. values between 0.07 -- 0.6 ${\rm M}_{\odot}$. If we apply this same approach to SN~1997bs, we find a $^{56}$Ni of 0.003 ${\rm M}_{\odot}$ based on the $V$--band magnitude measured 63.5 days after discovery.
In conclusion it is very difficult to set significant constraints on the nature of SN~2011A based on these uncertain $^{56}$Ni mass estimates.

\subsection{Spectroscopy}

\subsubsection{Explosion date}

Unfortunately, the explosion date of SN~2011A is not well determined by pre--explosion images. A fit to the early time spectrum with a blackbody function yields a high temperature ($\sim$ 10000 K), thus suggesting that the observations possibly began soon after explosion. However, given the interacting nature of SN~2011A, it is unclear how valid such a constraint is. To better estimate this parameter, we compare SN~2011A with two other SNe with well--constrained explosion times.\\
\indent

Firstly, SN~1994W \citep{cortini94}, for which \citet{sol98} estimated from its $R$--band data a precise explosion date of 1994 July $14_{-4}^{+2}$. In Fig.~14 we compare our spectrum taken 2.0 days after discovery with a SN~1994W spectrum taken 57 days after explosion. As we can see, the spectra are remarkably similar. A blue continuum characterizes both spectra and the slope/shape are nearly identical. Also, both SNe show strong \ion{Fe}{2} multiplet $\lambda\lambda 4923, 5018, 5169$ lines with low velocity ($\sim$ hundreds km s$^{-1}$) P--Cygni profiles (Fig.~14, top right). We also distinguish in the blue part of the spectrum a strong absorption attributed to CaII $\lambda\lambda 3933,3968$. SN~2011A shows two noticeable differences from SN~1994W: (1) the \ion{Ca}{2} near-infrared triplet is more prominent than in SN~1994W and (2) SN~1994W shows a narrow P--Cygni absorption on top of the broad of H$\alpha$ component whereas in SN~2011A this feature is not seen and only appears in the spectrum taken +56.8 days after discovery.\\
\indent

Secondly, in Fig.~15 we compare our +2.0 day spectrum with the SN~1997bs spectrum taken on 1997 April 16th (day +1 after discovery) to search for similarities as seen in their light curves. Given that the non--detection was very close to discovery, the explosion date is well constrained for SN~1997bs. The continua look very similar when we apply a total reddening of $E(B-V)$ = 0.24 mag to SN~1997bs (\citealt{van00} estimated this total color excess at 0.21 mag) and null for SN~2011A. We choose this colour excess value to match the two continua. The spectra show some differences, most notably the lack of narrow P--Cygni \ion{Fe}{2} lines in SN~1997bs. In contrast, their H$\alpha$ emission profiles are virtually identical. Note also that if these spectra were taken at similar epochs then this would imply that the host--galaxy extinction for SN~2011A is less than that of SN~1997bs. For SN~1997bs \citet{van00} derived an total extinction $A_{V}$ $\sim$ 0.65 mag. This value is close to that found for SN~2011A in section 3.1.3 but not significantly lower.\\

\indent
Thanks to these comparisons and the fact that our spectrum taken 2.0 days after discovery has a good match with the SN~1994W spectrum taken 57 days past explosion, it is possible that SN~2011A maybe exploded $\sim$ 50 days before discovery. Indeed in both spectra we can see the presence of the same elements with identical line profile, such as the \ion{Fe}{2} multiplet $\lambda\lambda 4923, 5018, 5169$. This implies that the physical conditions for both spectra were similar and assuming that the evolution is not so different, the epochs are likely similar. This explosion date would then make the colour curve between SN~2009kn and SN~2011A consistent (see section 4.1). While the light curves are very similar between SN~2011A and SN~1997bs these are most likely related to CSM shells, and therefore any explosion date estimation is less reliable; the explosion date estimation from spectral comparisons is likely to be more valid.

\begin{figure}
\epsscale{1.20}
\plotone{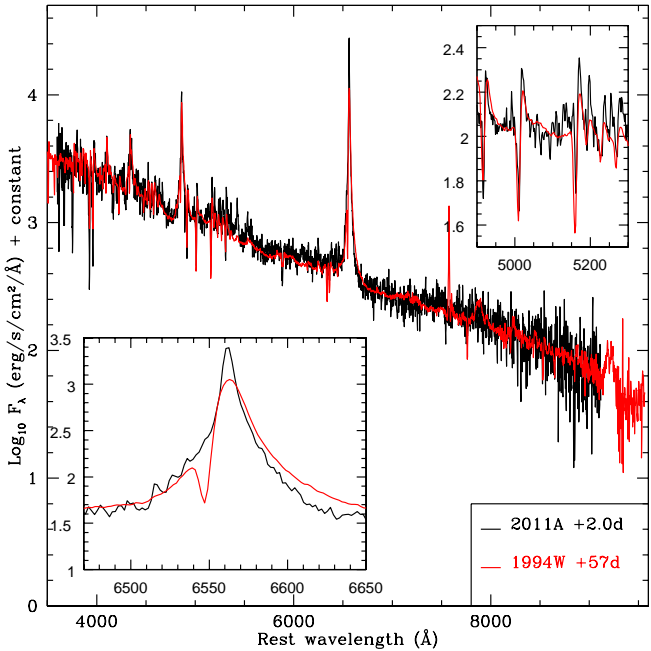}
\caption{Comparison between the SN~2011A spectrum taken 2.0 days after discovery, in black, and that of SN~1994W taken +57 days after explosion, in red. In the top right inset we zoom on of the \ion{Fe}{2} lines, $\lambda\lambda 4923, 5018, 5169$. On the bottom left inset we zoom on H$\alpha$ emission line. For both spectra we did not apply any MW or host--galaxy extinction correction.}
\label{Figure.14}
\end{figure}

\begin{figure}
\epsscale{1.2}
\plotone{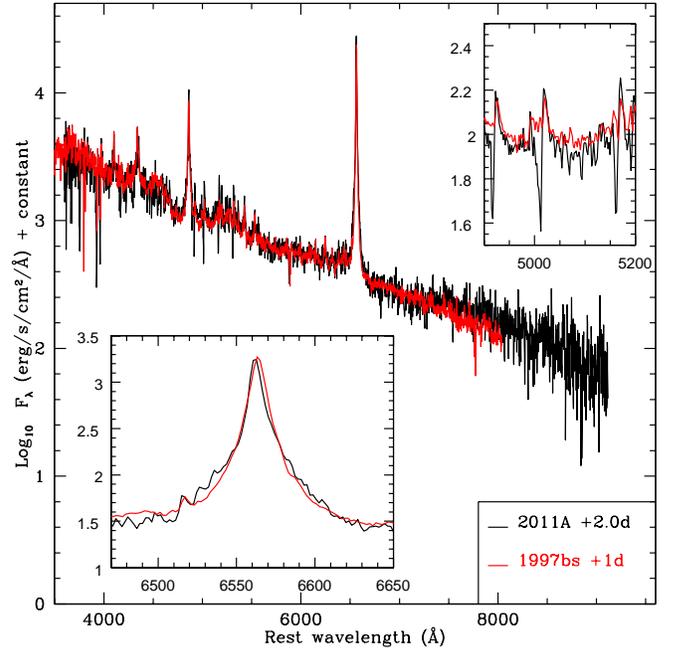}
\caption{Comparison between the SN~2011A spectrum taken 2.0 days after discovery, in black, and that of SN~1997bs taken on 16th April 1997, one day after discovery, in red (no detection on 10th Apr 1997). In the top right inset we zoom on the \ion{Fe}{2} lines, $\lambda\lambda 4923, 5018, 5169$. On the bottom left inset we zoom on the H$\alpha$ emission line. We applied a total reddening of $E(B-V)$=0.24 mag to SN~1997bs in order to match the continua.}
\label{Figure.15}
\end{figure}

\subsubsection{\ion{Na}{1} D evolution}

 \ion{Na}{1} D absorption can have three origins. From the interstellar medium (ISM) in the line of sight, the circumstellar material ejected by the progenitor ionized by the radiation from the SN and recombined over the following
several days to weeks, or from the SN ejecta. The main distinguishing feature between these three components is the velocity. If the absorption is due to the ISM, the velocity will be very low, $\sim$ 2 -- 15 km s$^{-1}$ \citep{fas01}. If the component comes from the CSM, the velocity will be higher (few hundred km s$^{-1}$ as for SN~2011A) and depends on the progenitor. Finally, the component from a SN ejecta would have the highest values of a few thousand  km s$^{-1}$.
Though the majority of spectra are low resolution and prevent us to resolve the \ion{Na}{1} D absorption line, we believe that the absorption close to the \ion{Na}{1} D wavelength position is mostly due to \ion{Na}{1} D. Even if some lines like \ion{Ba}{2} $\lambda 5854$ can be blended with the \ion{Na}{1} D, the absorption strength and velocity are very rare. Indeed others SNe have shown similar remarkable increase of \ion{Na}{1} D absorption, such as SN 1998A and SN~2009E \citep{pas05,pas12a}, but \citet{pas12} measured for SN~2009E the \ion{Na}{1} D velocity and found a value of $\sim$ 3000--5000 km s$^{-1}$. This implies that the absorption line is due the SN ejecta, whereas the velocities measured for \ion{Na}{1} D absorption in SN~2011A never gets higher than 1100 km s$^{-1}$, inconsistent with an ejecta origin, but consistent with a CSM origin interpretation.\\
\indent

In the literature we found a small number of cases of CSM \ion{Na}{1} D in SNe~IIn. Indeed for SN~2009kn \citep{kan12}, SN~1994W \citep{chu04}, SN~1997bs \citep{van00}, and SN~2005kj \citep{tad13}, the velocity is low as for SN~2011A, less than 1100 km s$^{-1}$. The increase of \ion{Na}{1} D absorption, according to \citet{chu94}, is a result of the increasing \ion{Na}{1} ionization fraction.\\ 

\subsubsection{Dust formation}
Most SN~impostors candidates in the literature have no surviving star at the location of the transient which could contradict the nature of the transient, i.e., a non--terminal eruptions of massive stars. However, a simple reason could explain this. The dust formed in a dense shell ejected by the star will obscures the SN~impostor. The presence of dust can be deduced by different ways. First, by the presence of an infrared excess due to the thermal emission from dust condensed in the ejecta or from heated pre--existing dust. Unfortunately we do not have useful infrared data for this transient. Secondly, looking at the emission lines. If the spectral feature profiles show a blueshift, this can be attributed to the fact that the red side of the ejecta is blocked by the new dust (\citealt{smith12}, but see \citealt{anderson14b} and \citealt{dessart15} for an alternative explanation). This effect is maybe seen in our transient. Indeed in Fig.~5 where the H$\alpha$ emission line evolution is shown, we can see for the spectrum +111.8 days that the red side is weaker compared to the blue part. This characteristic is not clearly observed after and before this epoch especially due to the appearance of a P--Cygni profile.\\
\indent

The estimated black body temperature at late times (3300K) is also possibly indicative of dust formation. This is because this temperature appears to be too low for the spectra to be dominated by strong Balmer emission lines. This low temperature could be due to dust formation which reddens the object. Multi-band photometry fitting seems to favor an extincted black body but it is difficult to put strong constraints on the black body temperature and extinction due to the small wavelength baseline and because our transient is mainly powered by interaction which implies that using a black body fitting could be not realistic.
Any increase in reddening could also related to the observed increase in \ion{Na}{1} D. However we note that the increase in the latter is more likely due to CSM effects.

\subsubsection{CSM properties} 

Analysing the H${\alpha}$ emission line evolution with time together with the light curve allows us to constrain CSM properties. The SN~2011A light curve shows a double plateau. The presence of this rare double plateau leads us to speculate that the CSM is composed of two shells ejected by the pre SN wind. During $\sim$ 15 days there is an initial plateau which corresponds to the interaction between the SN blast wave and the first shell. When the ejecta reaches the edge of the first shell the light curve drops. If we assume that the blast wave speed is constant and taking the value found in section 3.2.4, $\sim$ 2000 km s$^{-1}$, we find a lower limit width of the first shell of $\geq 2.6 \times 10^{14}$cm. It is a lower limit due to the fact the beginning of this shell is not well constrained (because of the uncertainty in the explosion date). Then, there is another weaker plateau which lasts $\sim$ 15 days in the $V$ band. Again, assuming that the ejecta velocity is constant with time, we find the width for the second shell, $\sim 2.6 \times 10^{14}$cm. A cartoon showing the CSM and the velocities of the different components is presented in Fig.~16. This scenario would be similar to that proposed by \citet{dessart09} for SN~1994W.\\
\indent

This photometric analysis seems be also consistent with the spectral evolution. In Fig.~5 we see that the H$\alpha$ profile changes with time, from the very broad component at early epochs during the first plateau phase (spectra +2.0, +2.4, and +14.9 days), to a narrower broad component, accompanied with low velocity H$\alpha$ absorption (+56.8, +64, +70, +80.9, +85.9 days) during and after the second plateau.

\begin{figure*}
\epsscale{1.1}
\plotone{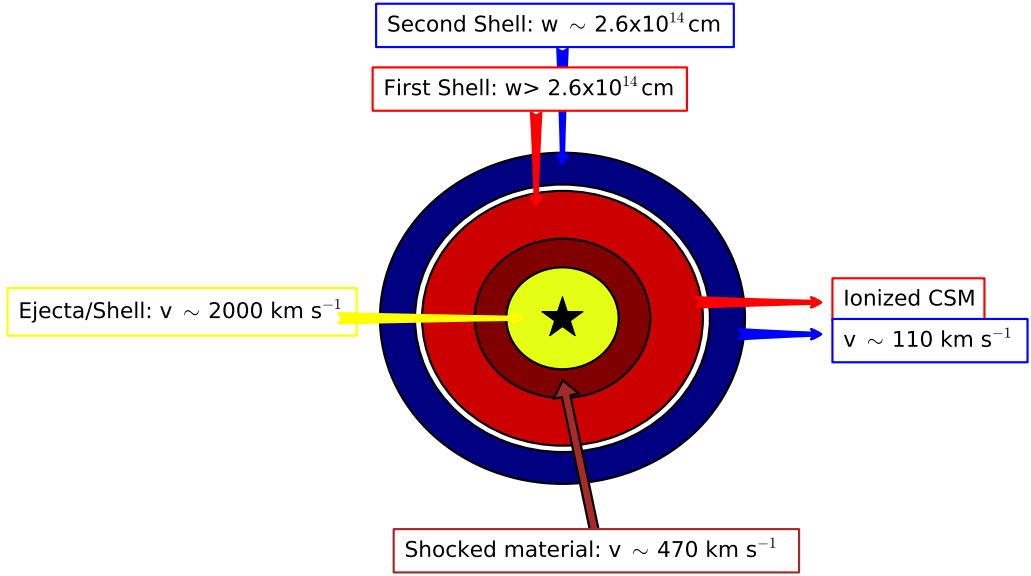}
\caption{Cartoon illustration of the CSM composition with the velocity of the different components. The black star represents the transient. Not to scale.}
\label{Figure.16}
\end{figure*}

\section{Conclusions}

We have presented  a detailed study of SN~2011A, initially classified as a SN~IIn based on the presence of a narrow H$\alpha$ emission in the spectrum. While the exact nature of the transient remains unclear, SN~2011A shows a number of interesting and unusual properties:
\begin{enumerate}
\item{ Double plateau in light curve likely due to CSM composed of two shells.}
\item{ Low luminosity, $-$15.10$~\geq$ $M_{V}$ $\geq$ $-$15.72 (depending on assumed host--galaxy extinction, 0 or 0.62 mag in $V$--band).}
\item{ Low P--Cygni H$\alpha$ velocity $\leq 1200$ km s$^{-1}$.}
\item{ Low velocity absorption close to \ion{Na}{1} D $\leq 1100$ km s$^{-1}$.}
\item{ Variable absorption close to \ion{Na}{1} D increasing in strength, from 3 to 10 \AA ~in equivalent width.}
\end{enumerate}

Given the photometric and spectroscopic properties we have derived for SN~2011A; most importantly its low luminosity and low ejecta velocity together with their comparison to other interacting transients, we believe that SN~2011A is most likely an SN~impostor event. However, we also note that our analysis has demonstrated the difficulty in determining whether interacting transients are true terminal events or indeed SN~impostors. The reader also must keep in mind that the characteristics used to separate the two events are in general arbitrary and hence detailed studies of future events are needed to further elucidate this issue.\\
\\
\section{Acknowledgements}
\acknowledgments
The referee is thanked for their through reading of the manuscript, which helped clarify and improve the paper. Support for TdJ, GP, MH, FF, CG, FB, and SG is provided by the Ministry of Economy, Development, and Tourism's Millennium Science Initiative through grant IC120009, awarded to The Millennium Institute of Astrophysics, MAS. S.G., and F.B. acknowledges support by CONICYT through FONDECYT grants 3130680 and 3120227. SB and LT are partially supported by the PRIN--INAF 2011 with the project ``Transient Universe: from ESO Large to PESSTO''. N.E.R. acknowledges financial support by the MICINN grant AYA2011--24704/ESP, and by the ESF EUROCORES Program EuroGENESIS (MINECO grants EUI2009--04170). M. S. gratefully acknowledge generous support provided by the Danish Agency for Science and Technology and Innovation realized through a Sapere Aude Level 2 grant. E.K. acknowledges financial support from the Jenny and Wihuri Foundation. M. D. S. gratefully acknowledges generous support provided by the Danish Agency for Science and Technology and Innovation realized through a Sapere Aude Level 2 grant. The authors thank J. Mauerhan for providing us with the SN~1994W spectrum (taken from the A. Filippenko's database in Berkeley) and also S. D. Van Dyk for the SN~1997bs spectrum.
This research has made use of the NASA/IPAC Extragalactic Database (NED) which is operated by the Jet Propulsion Laboratory, California Institute of Technology, under contract with the National Aeronautics and Space Administration and of data provided by the Central Bureau for Astronomical Telegrams.


\begin{thebibliography}{95}
\expandafter\ifx\csname natexlab\endcsname\relax\def\natexlab#1{#1}\fi

\bibitem[{{Abbott} \& {Conti}(1987)}]{abb87}
{Abbott}, D.~C., \& {Conti}, P.~S. 1987, \araa, 25, 113

\bibitem[{{Adams} \& {Kochanek}(2015)}]{adams15}
{Adams}, S.~M., \& {Kochanek}, C.~S. 2015, ArXiv e-prints

\bibitem[{{Anderson} {et~al.}(2012){Anderson}, {Habergham}, {James}, \&
  {Hamuy}}]{and12}
{Anderson}, J.~P., {Habergham}, S.~M., {James}, P.~A., \& {Hamuy}, M. 2012,
  \mnras, 424, 1372

\bibitem[{{Anderson} {et~al.}(2014{\natexlab{a}})}]{anderson14b}
{Anderson}, J.~P., {et~al.} 2014{\natexlab{a}}, \mnras, 441, 671

\bibitem[{{Anderson} {et~al.}(2014{\natexlab{b}})}]{and14}
---. 2014{\natexlab{b}}, \apj, 786, 67

\bibitem[{{Arcavi} {et~al.}(2011)}]{arcavi11}
{Arcavi}, I., {et~al.} 2011, \apjl, 742, L18

\bibitem[{{Bond} {et~al.}(2009)}]{bond09}
{Bond}, H.~E., {et~al.} 2009, \apjl, 695, L154

\bibitem[{{Botticella} {et~al.}(2009)}]{bot2009}
{Botticella}, M.~T., {et~al.} 2009, \mnras, 398, 1041

\bibitem[{{Buzzoni} {et~al.}(1984){Buzzoni}, {Delabre}, {Dekker}, {Dodorico},
  {Enard}, {Focardi}, {Gustafsson}, {Nees}, {Paureau}, \&
  {Reiss}}]{buzzoni1984}
{Buzzoni}, B., {et~al.} 1984, The Messenger, 38, 9

\bibitem[{{Campana} {et~al.}(2006)}]{campana06}
{Campana}, S., {et~al.} 2006, \nat, 442, 1008

\bibitem[{{Cardelli} {et~al.}(1989){Cardelli}, {Clayton}, \& {Mathis}}]{car89}
{Cardelli}, J.~A., {Clayton}, G.~C., \& {Mathis}, J.~S. 1989, \apj, 345, 245

\bibitem[{{Chevalier}(1981)}]{che81}
{Chevalier}, R.~A. 1981, \apj, 251, 259

\bibitem[{{Chugai}(1990)}]{chu90}
{Chugai}, N.~N. 1990, Soviet Astronomy Letters, 16, 457

\bibitem[{{Chugai}(2001)}]{chu01}
---. 2001, \mnras, 326, 1448

\bibitem[{{Chugai} \& {Danziger}(1994)}]{chu94}
{Chugai}, N.~N., \& {Danziger}, I.~J. 1994, \mnras, 268, 173

\bibitem[{{Chugai} {et~al.}(2004)}]{chu04}
{Chugai}, N.~N., {et~al.} 2004, \mnras, 352, 1213

\bibitem[{{Cortini} {et~al.}(1994)}]{cortini94}
{Cortini}, G., {et~al.} 1994, \iaucirc, 6042, 1

\bibitem[{{Crowther}(2007)}]{crowther2007}
{Crowther}, P.~A. 2007, \araa, 45, 177

\bibitem[{{Dessart} {et~al.}(2015){Dessart}, {Audit}, \& {Hillier}}]{dessart15}
{Dessart}, L., {Audit}, E., \& {Hillier}, D.~J. 2015, ArXiv e-prints

\bibitem[{{Dessart} {et~al.}(2009){Dessart}, {Hillier}, {Gezari}, {Basa}, \&
  {Matheson}}]{dessart09}
{Dessart}, L., {Hillier}, D.~J., {Gezari}, S., {Basa}, S., \& {Matheson}, T.
  2009, \mnras, 394, 21

\bibitem[{{Dilday} {et~al.}(2012)}]{dil12}
{Dilday}, {et~al.} 2012, Science, 337, 942

\bibitem[{{Dwarkadas} {et~al.}(2010){Dwarkadas}, {Dewey}, \& {Bauer}}]{dwa10}
{Dwarkadas}, V.~V., {Dewey}, D., \& {Bauer}, F. 2010, \mnras, 407, 812

\bibitem[{{Faran} {et~al.}(2014)}]{faran14}
{Faran}, T., {et~al.} 2014, \mnras, 442, 844

\bibitem[{{Fassia} {et~al.}(2001)}]{fas01}
{Fassia}, {et~al.} 2001, \mnras, 325, 907

\bibitem[{{Filippenko}(1997)}]{filippenko97}
{Filippenko}, A.~V. 1997, \araa, 35, 309

\bibitem[{{Fransson}(1982)}]{fra82}
{Fransson}, C. 1982, \aap, 111, 140

\bibitem[{{Fraser} {et~al.}(2013)}]{fraser13}
{Fraser}, M., {et~al.} 2013, \mnras, 433, 1312

\bibitem[{{Gal-Yam} \& {Leonard}(2009)}]{gal09}
{Gal-Yam}, A., \& {Leonard}, D.~C. 2009, \nat, 458, 865

\bibitem[{{Gr{\"a}fener} \& {Hamann}(2008)}]{gra08}
{Gr{\"a}fener}, G., \& {Hamann}, W.-R. 2008, \aap, 482, 945

\bibitem[{{Groh} {et~al.}(2013){Groh}, {Meynet}, \& {Ekstr{\"o}m}}]{groh2013}
{Groh}, J.~H., {Meynet}, G., \& {Ekstr{\"o}m}, S. 2013, \aap, 550, L7

\bibitem[{{Habergham} {et~al.}(2014)}]{habergham14}
{Habergham}, S.~M., {et~al.} 2014, \mnras, 441, 2230

\bibitem[{{Hamuy}(2003)}]{ham03a}
{Hamuy}, M. 2003, \apj, 582, 905

\bibitem[{{Hamuy} {et~al.}(1994){Hamuy}, {Suntzeff}, {Heathcote}, {Walker},
  {Gigoux}, \& {Phillips}}]{ham94}
{Hamuy}, M., {Suntzeff}, N.~B., {Heathcote}, S.~R., {Walker}, A.~R., {Gigoux},
  P., \& {Phillips}, M.~M. 1994, \pasp, 106, 566

\bibitem[{{Hamuy} {et~al.}(1992){Hamuy}, {Walker}, {Suntzeff}, {Gigoux},
  {Heathcote}, \& {Phillips}}]{ham92}
{Hamuy}, M., {Walker}, A.~R., {Suntzeff}, N.~B., {Gigoux}, P., {Heathcote},
  S.~R., \& {Phillips}, M.~M. 1992, \pasp, 104, 533

\bibitem[{{Hamuy} {et~al.}(2003)}]{hamuy03}
{Hamuy}, M., {et~al.} 2003, \nat, 424, 651

\bibitem[{{Hamuy} {et~al.}(2006)}]{ham06}
---. 2006, \pasp, 118, 2

\bibitem[{{Humphreys} \& {Davidson}(1994)}]{hum94}
{Humphreys}, R.~M., \& {Davidson}, K. 1994, \pasp, 106, 1025

\bibitem[{{Kankare} {et~al.}(2012)}]{kan12}
{Kankare}, {et~al.} 2012, \mnras, 424, 855

\bibitem[{{Kiewe} {et~al.}(2012)}]{kie12}
{Kiewe}, M., {et~al.} 2012, \apj, 744, 10

\bibitem[{{Kochanek}(2011)}]{kochanek11}
{Kochanek}, C.~S. 2011, \apj, 741, 37

\bibitem[{{Landolt}(1992)}]{lan92}
{Landolt}, A.~U. 1992, \aj, 104, 340

\bibitem[{{Landolt}(2007)}]{lan07}
---. 2007, \aj, 133, 2502

\bibitem[{{Langer}(1993)}]{lan93}
{Langer}, N. 1993, \ssr, 66, 365

\bibitem[{{Li} {et~al.}(2002){Li}, {Filippenko}, {Van Dyk}, {Hu}, {Qiu},
  {Modjaz}, \& {Leonard}}]{li2002}
{Li}, W., {Filippenko}, A.~V., {Van Dyk}, S.~D., {Hu}, J., {Qiu}, Y., {Modjaz},
  M., \& {Leonard}, D.~C. 2002, \pasp, 114, 403

\bibitem[{{Maeder} \& {Meynet}(2008)}]{mae08}
{Maeder}, A., \& {Meynet}, G. 2008, in Astronomical Society of the Pacific
  Conference Series, Vol. 388, Mass Loss from Stars and the Evolution of
  Stellar Clusters, ed. A.~{de Koter}, L.~J. {Smith}, \& L.~B.~F.~M. {Waters},
  3

\bibitem[{{Maeder} {et~al.}(2005){Maeder}, {Meynet}, \& {Hirschi}}]{mae05}
{Maeder}, A., {Meynet}, G., \& {Hirschi}, R. 2005, in Astronomical Society of
  the Pacific Conference Series, Vol. 336, Cosmic Abundances as Records of
  Stellar Evolution and Nucleosynthesis, ed. T.~G. {Barnes}, III \& F.~N.
  {Bash}, 79

\bibitem[{{Margutti} {et~al.}(2014)}]{margutti13}
{Margutti}, R., {et~al.} 2014, \apj, 780, 21

\bibitem[{{Mauerhan} {et~al.}(2013){Mauerhan}, {Smith}, {Filippenko},
  {Silverman}, {Cenko}, \& {Clubb}}]{mau12}
{Mauerhan}, J., {Smith}, N., {Filippenko}, A.~V., {Silverman}, J., {Cenko}, B.,
  \& {Clubb}, K. 2013, in American Astronomical Society Meeting Abstracts, Vol.
  221, American Astronomical Society Meeting Abstracts, 233.03

\bibitem[{{Maund} {et~al.}(2006)}]{maund06}
{Maund}, J.~R., {et~al.} 2006, \mnras, 369, 390

\bibitem[{{Mazzali} {et~al.}(2009){Mazzali}, {Deng}, {Hamuy}, \&
  {Nomoto}}]{mazzali09}
{Mazzali}, P.~A., {Deng}, J., {Hamuy}, M., \& {Nomoto}, K. 2009, \apj, 703,
  1624

\bibitem[{{Moriya} {et~al.}(2014){Moriya}, {Maeda}, {Taddia}, {Sollerman},
  {Blinnikov}, \& {Sorokina}}]{moriya14}
{Moriya}, T.~J., {Maeda}, K., {Taddia}, F., {Sollerman}, J., {Blinnikov},
  S.~I., \& {Sorokina}, E.~I. 2014, \mnras, 439, 2917

\bibitem[{{Nakar} \& {Piro}(2014)}]{nakar14}
{Nakar}, E., \& {Piro}, A.~L. 2014, \apj, 788, 193

\bibitem[{{Pastorello} {et~al.}(2004)}]{pas04}
{Pastorello}, A., {et~al.} 2004, \mnras, 347, 74

\bibitem[{{Pastorello} {et~al.}(2005)}]{pas05}
---. 2005, \mnras, 360, 950

\bibitem[{{Pastorello} {et~al.}(2010)}]{pastorello2010}
---. 2010, \mnras, 408, 181

\bibitem[{{Pastorello} {et~al.}(2012{\natexlab{a}})}]{pas12}
---. 2012{\natexlab{a}}, ArXiv e-prints

\bibitem[{{Pastorello} {et~al.}(2012{\natexlab{b}})}]{pas12a}
---. 2012{\natexlab{b}}, \aap, 537, A141

\bibitem[{{Patat}(1996)}]{patat1996}
{Patat}, F. 1996, PhD thesis, Univ. Padova

\bibitem[{{Phillips} {et~al.}(2013)}]{phillips13}
{Phillips}, M.~M., {et~al.} 2013, \apj, 779, 38

\bibitem[{{Pignata} {et~al.}(2009)}]{pig09}
{Pignata}, G., {et~al.} 2009, in American Institute of Physics Conference
  Series, Vol. 1111, American Institute of Physics Conference Series, ed.
  G.~{Giobbi}, A.~{Tornambe}, G.~{Raimondo}, M.~{Limongi}, L.~A. {Antonelli},
  N.~{Menci}, \& E.~{Brocato}, 551--554

\bibitem[{{Pignata} {et~al.}(2011)}]{pig11}
{Pignata}, G., {et~al.} 2011, Central Bureau Electronic Telegrams, 2623, 1

\bibitem[{{Prieto} {et~al.}(2013){Prieto}, {Brimacombe}, {Drake}, \&
  {Howerton}}]{pri12}
{Prieto}, J.~L., {Brimacombe}, J., {Drake}, A.~J., \& {Howerton}, S. 2013,
  \apjl, 763, L27

\bibitem[{{Prieto} {et~al.}(2009)}]{prieto09}
{Prieto}, J.~L., {et~al.} 2009, \apj, 705, 1425

\bibitem[{{Reichart} {et~al.}(2005)}]{reichart05}
{Reichart}, D., {et~al.} 2005, Nuovo Cimento C Geophysics Space Physics C, 28,
  767

\bibitem[{{Richardson} {et~al.}(2002)}]{richardson02}
{Richardson}, D., {et~al.} 2002, \aj, 123, 745

\bibitem[{{Richmond} {et~al.}(1994)}]{richmond94}
{Richmond}, M.~W., {et~al.} 1994, \aj, 107, 1022

\bibitem[{{Salamanca} {et~al.}(1998){Salamanca}, {Cid-Fernandes},
  {Tenorio-Tagle}, {Telles}, {Terlevich}, \& {Munoz-Tunon}}]{sal98}
{Salamanca}, I., {Cid-Fernandes}, R., {Tenorio-Tagle}, G., {Telles}, E.,
  {Terlevich}, R.~J., \& {Munoz-Tunon}, C. 1998, \mnras, 300, L17

\bibitem[{{Schaller} {et~al.}(1992){Schaller}, {Schaerer}, {Meynet}, \&
  {Maeder}}]{sch92}
{Schaller}, G., {Schaerer}, D., {Meynet}, G., \& {Maeder}, A. 1992, \aaps, 96,
  269

\bibitem[{{Schlafly} \& {Finkbeiner}(2011)}]{schlafly11}
{Schlafly}, E.~F., \& {Finkbeiner}, D.~P. 2011, \apj, 737, 103

\bibitem[{{Schlegel}(1990)}]{sch90}
{Schlegel}, E.~M. 1990, \mnras, 244, 269

\bibitem[{{Smartt}(2009)}]{smart09b}
{Smartt}, S.~J. 2009, \araa, 47, 63

\bibitem[{{Smartt}(2015)}]{smartt15}
---. 2015, ArXiv e-prints

\bibitem[{{Smith} {et~al.}(2002)}]{smithja2002}
{Smith}, J.~A., {et~al.} 2002, \aj, 123, 2121

\bibitem[{{Smith}(2013)}]{smi13}
{Smith}, N. 2013, \mnras, 434, 102

\bibitem[{{Smith} {et~al.}(2014){Smith}, {Mauerhan}, \& {Prieto}}]{smith14b}
{Smith}, N., {Mauerhan}, J.~C., \& {Prieto}, J.~L. 2014, \mnras, 438, 1191

\bibitem[{{Smith} \& {Tombleson}(2015)}]{smith15}
{Smith}, N., \& {Tombleson}, R. 2015, \mnras, 447, 598

\bibitem[{{Smith} \& {Townsend}(2007)}]{smi07a}
{Smith}, N., \& {Townsend}, R.~H.~D. 2007, \apj, 666, 967

\bibitem[{{Smith} {et~al.}(2007)}]{smi07b}
{Smith}, N., {et~al.} 2007, \apj, 666, 1116

\bibitem[{{Smith} {et~al.}(2009)}]{smi09}
---. 2009, \apjl, 697, L49

\bibitem[{{Smith} {et~al.}(2011)}]{smith11}
---. 2011, \mnras, 415, 773

\bibitem[{{Smith} {et~al.}(2012)}]{smith12}
---. 2012, \aj, 143, 17

\bibitem[{{Sollerman} {et~al.}(1998){Sollerman}, {Cumming}, \&
  {Lundqvist}}]{sol98}
{Sollerman}, J., {Cumming}, R.~J., \& {Lundqvist}, P. 1998, \apj, 493, 933

\bibitem[{{Steele} {et~al.}(2004)}]{steele2004}
{Steele}, I.~A., {et~al.} 2004, in Society of Photo-Optical Instrumentation
  Engineers (SPIE) Conference Series, Vol. 5489, Society of Photo-Optical
  Instrumentation Engineers (SPIE) Conference Series, ed. J.~M. {Oschmann},
  Jr., 679--692

\bibitem[{{Stothers} \& {Chin}(1996)}]{sto96}
{Stothers}, R.~B., \& {Chin}, C.-W. 1996, \apj, 468, 842

\bibitem[{{Stritzinger} {et~al.}(2011){Stritzinger}, {Prieto}, {Morrell}, \&
  {Pignata}}]{str11}
{Stritzinger}, M., {Prieto}, J.~L., {Morrell}, N., \& {Pignata}, G. 2011,
  Central Bureau Electronic Telegrams, 2623, 2

\bibitem[{{Taddia} {et~al.}(2013)}]{tad13}
{Taddia}, F., {et~al.} 2013, \aap, 555, A10

\bibitem[{{Tartaglia} {et~al.}(2014)}]{tartaglia14}
{Tartaglia}, L., {et~al.} 2014, ArXiv e-prints

\bibitem[{{Theureau} {et~al.}(2007){Theureau}, {Hanski}, {Coudreau}, {Hallet},
  \& {Martin}}]{theureau07}
{Theureau}, G., {Hanski}, M.~O., {Coudreau}, N., {Hallet}, N., \& {Martin},
  J.-M. 2007, \aap, 465, 71

\bibitem[{{Umeda} \& {Nomoto}(2008)}]{ume08}
{Umeda}, H., \& {Nomoto}, K. 2008, \apj, 673, 1014

\bibitem[{{Van Dyk} {et~al.}(2000){Van Dyk}, {Peng}, {King}, {Filippenko},
  {Treffers}, {Li}, \& {Richmond}}]{van00}
{Van Dyk}, S.~D., {Peng}, C.~Y., {King}, J.~Y., {Filippenko}, A.~V.,
  {Treffers}, R.~R., {Li}, W., \& {Richmond}, M.~W. 2000, \pasp, 112, 1532

\bibitem[{{Van Dyk} {et~al.}(2006)}]{van06}
{Van Dyk}, S.~D., {et~al.} 2006, ArXiv Astrophysics e-prints

\bibitem[{{Vink}(2011)}]{vin11}
{Vink}, J.~S. 2011, \apss, 336, 163

\bibitem[{{Wagner} {et~al.}(2004)}]{wagner2004}
{Wagner}, R.~M., {et~al.} 2004, \pasp, 116, 326

\bibitem[{{Weis} \& {Bomans}(2005)}]{weis05}
{Weis}, K., \& {Bomans}, D.~J. 2005, \aap, 429, L13

\bibitem[{{Zhang} {et~al.}(2012)}]{zhang12}
{Zhang}, T., {et~al.} 2012, \aj, 144, 131

\end{thebibliography}
\end{document}